\documentclass[english]{article}
\usepackage{siunitx}
\usepackage{mathrsfs}
\usepackage{bm}
\usepackage{amsmath}
\usepackage{amsthm}
\usepackage{amssymb}
\usepackage{graphicx}
\usepackage{esint}

\makeatletter
\theoremstyle{plain}

\theoremstyle{plain}

\ifx\proof\undefined
\newenvironment{proof}[1][\protect\proofname]{\par
	\normalfont\topsep6\p@\@plus6\p@\relax
	\trivlist
	\itemindent\parindent
	\item[\hskip\labelsep\scshape #1]\ignorespaces
}{%
	\endtrivlist\@endpefalse
}
\providecommand{\proofname}{Proof}
\fi

\usepackage{jhepmod}
\pdfoutput=1

\usepackage[table ]{ xcolor}

\usepackage{amssymb}
\usepackage{graphicx,color}

\usepackage{amsmath,amsthm}

\usepackage{mathrsfs}

\usepackage{bm}

\def\beq{\begin{equation}}
\def\eeq{\end{equation}}
\newcommand{\bea}{\begin{eqnarray}}
\newcommand{\eea}{\end{eqnarray}}
\def\bi{\begin{itemize}}
\def\ei{\end{itemize}}
\def\ba{\begin{array}}
\def\ea{\end{array}}
\def\bfig{\begin{figure}}
\def\efig{\end{figure}}

\def\C{\mathbb{C}}

\newcommand{\Slc}{\mathrm{SL}(2,\mathbb{C})}

\newcommand{\Su}{\mathrm{SU}(2)}

\def\be{\begin{eqnarray}}
\def\ee{\end{eqnarray}}

\newcommand{\ch}{\mathcal H}

\newcommand{\cn}{\mathcal N}

\newcommand{\cv}{\mathcal V}

\newcommand{\cx}{\mathcal X}
\newcommand{\cy}{\mathcal Y}

\renewcommand{\a}{\alpha}
\renewcommand{\b}{\beta}
\newcommand{\g}{\gamma}
\newcommand{\G}{\Gamma}

\renewcommand{\l}{\lambda}

\renewcommand{\t}{\tau}

\newcommand{\rmd}{\mathrm d}

\newcommand{\lt}{\left}
\newcommand{\rt}{\right}

\newcommand{\im}{\mathrm{Im}}

\title{The one-loop effective action from the coherent state path integral of loop quantum gravity}

\author[1]{Renata Ferrero}
\emailAdd{renata.ferrero(AT)gravity.fau.de}

\author[1,2]{\ Muxin Han} \emailAdd{hanm(AT)fau.edu}

\author[3,4,1]{\ Hongguang Liu} \emailAdd{liuhongguang(AT)westlake.edu.cn}

\affiliation[1]{Department Physik, Institut f\"ur Quantengravitation, Theoretische Physik III, Friedrich-Alexander Universit\"at Erlangen-N\"urnberg, Staudtstr. 7/B2, 91058 Erlangen, Germany}

\affiliation[2]{Department of Physics, Florida Atlantic University, 777 Glades Road, Boca Raton, FL 33431-0991, USA}
\affiliation[3]{Institute for Theoretical Sciences, Westlake University, Hangzhou 310030, China}
\affiliation[4]{Institute of Natural Sciences, Westlake Institute for Advanced Study, Hangzhou 310024, China}

\abstract{ 
We adopt a novel approach to combine path integral methods  with Loop Quantum Gravity (LQG). Our approach builds upon the recently developed coherent state path integral formulation of LQG to compute the one-loop effective action. We compare this methodology with the conventional Quantum Field Theory (QFT) prescription for path integrals and extend the formalism to account for the dependence on  boundary (coherent) states. This work aims to explore two aspects: to compare our results with the divergences observed in one-loop calculations of Einstein gravity testing UV-finiteness  and to initiate an exploration of the IR effective properties of LQG. We compute the effective action around flat spacetime obtaining  analytical and numerical results in the long and short wavelength approximations, respectively. Due to the one-loop dynamics of the LQG area, we find a divergence-free effective action.  We study the propagator and the dynamical modes and derive the quantum equation of motion at one loop. We ensure consistency with the semiclassical approximation in the long wavelength limit and, beyond this approximation, analyze numerical scaling as the lattice size increases.
}
\begin{document}

\maketitle

\section{Introduction}
In the last decades two very successful theories have been developed: Quantum Field Theory (QFT), which explains phenomena at sub-atomic scales, and General Relativity (GR), which governs interactions at macroscopic to cosmic scales. Both are effective within their domains and align with current experiments \cite{LIGOScientific:2016lio,Carney:2018ofe,ParticleDataGroup:2024cfk}. However, they rest on fundamentally different principles and mathematical frameworks, suggesting the need for a unified theory that incorporates gravity into quantum mechanics for a more comprehensive understanding of all interactions. 
Noteworthy approaches include adapting QFT path integral methods to accommodate gravitational interactions (or the sum over geometries in the discrete counterpart) on one side, and attempts to directly quantizing the gravitational Hamiltonian on the other side. The first category includes %
the Asymptotic Safety (AS) approach, which redefines renormalizability through  non-perturbative renormalization group (RG) techniques (see \cite{Percacci:2017fkn, Reuter:2019byg,Bonanno:2020bil,Saueressig:2023irs} for the continuum and \cite{Ambjorn:2012jv,Loll:2019rdj,Ambjorn:2024qoe,Ambjorn:2024bud} for the discrete implementation). In the second category, the most prominent example is the canonical formulation of Loop Quantum Gravity (LQG), which introduces new variables, such as holonomies and fluxes, to quantize the gravitational interaction in a discrete and non-perturbative fashion \cite{Rovelli:2004tv,Thiemann:2007pyv,rovelli2014covariant,Thiemann:2023zjd,Ashtekar:2021kfp}. From LQG various covariant path integral approaches have been developed, such as spin foams \cite{rovelli2014covariant,Oriti:2001qu,Perez:2003vx,Engle:2023qsu,Han:2021tzw,Han:2021kll}, effective spin foams \cite{Asante:2020qpa,Asante:2022dnj,Borissova:2022clg,Borissova:2024txs} and Group Field Theory \cite{Freidel:2005qe,Oriti:2013aqa,Marchetti:2024nnk,Marchetti:2024tjq}. Both AS and LQG are  non-perturbative approaches to quantum gravity and they both comply with the principle of background independence, even though this is realized in a different manner \cite{Ashtekar:2004eh,Ashtekar:2014kba,Thiemann:2024vjx}.

The lesson we learn from Quantum Field Theory applied to gravity is that some good properties observed in the low energy regime can become potential risks when dealing with high energies \cite{Goroff:1985sz, Goroff:1985th} and the specifics of the microscopic theory.
On the other hand, on a more pragmatical footing, this boils down to the problem of constructing infrared (IR) theoretical models from unknown ultraviolet (UV) degrees of freedom. Tackling this problem is the target of the \textit{effective approach}. Its core principle consists in constructing IR theories  using a combination of renormalization methods and power expansions of a suitable cutoff \cite{Peskin:1995ev,Weinberg:1996kr,Toms:2012bra,Schwartz:2014sze}. This should ultimately lead to select which of the possible bare theories correctly reproduces the IR physics (we refer to \cite{Eichhorn:2024rkc} for a recent overview in a quantum gravity context).

The major challenge arises when attempting to find a UV complete theory, meaning a theory that works at all energy levels \cite{Weinberg:1980gg} and remains valid as the cutoff is removed (for an UV cutoff, sent to infinity). A UV complete theory  renders a theory predictive and can be potentially matched with experiments.

As  it is well known, in the case of Einstein gravity the effective action presents ultraviolet and infrared divergences \cite{Birrell:1982ix,Parker:2009uva,Percacci:2017fkn}. Consequently, a common procedure is to readjust each bare divergent contribution (for instance infinities appearing in Feynman diagrams) by a suitable counterterm to give a ``renormalized'' quantity: this was the early stage of the perturbative renormalization procedure. The numerical values of the renormalized quantities have to be determined by experiments. However, in quantized General Relativity it turns out that there should be infinitely many such renormalized parameters: gravity is perturbatively non-renormalizable at two-loops \cite{Goroff:1985sz, Goroff:1985th,vandeVen:1991gw}.

An alternative was developed within the Asymptotic Safety scenario. Inspired by the Wilsonian Renormalization Group (RG) \cite{Wilson:1973jj,Kadanoff:2000xz}, this approach takes an indirect route by reconstructing the functional integral in the continuum limit through the solution of a suitable Functional Renormalization Group (FRG) equation \cite{Saueressig:2023irs}. This idea underpins the AS scenario, where the UV completion of quantum gravity is achieved through a non-trivial fixed point in the FRG flow. This scenario was first pointed out by Weinberg in 1979 \cite{Weinberg:1980gg}, concretely implemented by Reuter in 1996 \cite{Reuter:1996cp} and recently revised  by several authors within a suitable perturbative treatment of the path integral \cite{Niedermaier:2010zz,Martini:2021slj,Kluth:2024lar,Falls:2024noj}. Importantly, in those computations, the background is kept arbitrary throughout the calculations via the so-called background field method, preserving the principle of background independence \cite{DeWitt:1964mxt,DeWitt:1975ys}.
Therefore, one might currently question whether the problematic results of directly applying standard QFT tools to quantum gravity stem more from a lack of background independence rather than from the perturbative nature of the methods (see \cite{Thiemann:2024tmv} for a discussion about the interplay between background independence and (non-)perturbativity). 

It has to be acknowledged, that QFT has been very successful in matching high energy precision measurements  in particle physics \cite{ParticleDataGroup:2024cfk}. Furthermore, particularly for our purposes, the main advantage of QFT approaches is that they provide tools to directly access the effective IR theory, where all degrees of freedom have been integrated out, starting from a UV complete theory.
On the other hand, a QFT treatment might be lacking a full quantization. In particular, the influence of the choice of states, which is crucial in a Lorentzian setting \cite{Banerjee:2022xvi,DAngelo:2022vsh,DAngelo:2023wje,DAngelo:2025yoy,Thiemann:2024vjx}, and physical degrees of freedom can  not be explicitly kept track of, causing related properties to be lost along the computations.

Crucially, these are precisely the properties that the LQG approach in its canonical formalism is equipped to address.  In particular, in the reduced phase space quantization of LQG \cite{Thiemann:2004wk,Giesel:2007wn,Han:2009aw,Giesel:2012rb,Thiemann:2023zjd} one can study the quantum dynamics given by the full LQG physical Hamiltonian defined on the physical Hilbert space
of the physical observables. This procedure is based on classical deparametrized models \cite{Dittrich:2004cb} of gravity such as gravity coupled to scalar fields or  dust \cite{Brown:1994py} and has found fruitful applications both in cosmology \cite{Giesel:2007wi,Giesel:2007wk,Giesel:2020raf,Han:2020iwk,Han:2019feb, Han:2021cwb} and black hole physics \cite{Giesel:2022rxi,Thiemann:2024nmy,Neuser:2024bpq}.

An important advancement towards the application of AS techniques to a deparametrized physical Hamiltonian via the construction of a phase space path integral has been performed in \cite{Thiemann:2024vjx,Ferrero:2024rvi}. In this work we will take a different complementary approach: we will exploit the recently developed coherent state path integral of LQG \cite{Han:2019vpw} to construct the effective dynamics by means of the derivation of the one-loop effective action.

It has been shown \cite{Thiemann:2000bw,Thiemann:2000ca,Thiemann:2000bx}  that the Hilbert space of physical degrees of freedom possesses an overcomplete basis formed by the complexifier coherent states. A coherent state label serves as the holomorphic coordinate on the LQG phase space, parameterized by the holonomies and the gauge covariant fluxes. Due to the overcompleteness and semiclassical properties of coherent states, along with the standard discretization and coherent state path integral approach, the transition amplitude between gauge-invariant coherent states can be expressed as a discrete path integral formula \cite{Han:2019vpw,Han:2020iwk, Han:2021cwb}. This new formulation has shed new light on the link between covariant and the canonical formalism and can furnish insights about the embedding of symmetry reduced models, within the full theory, such as for Loop Quantum Cosmology (LQC) \cite{Agullo:2016tjh}.

As a first investigation, in this paper we well use a flat background and compute the one-loop functional determinant of the coherent state path integral. This comes with the advantage that we can access the Fourier space.

The purpose of this work is twofold: on one hand, we express, for the first time, the high-energy properties of LQG using the language of QFT divergences; on the other hand, we compute the one-loop effective action, enabling us to investigate the effective on-shell dynamics of LQG.

As an important perspective, our work suggests that the effective theory from LQG should be free of UV divergences. Let us first illustrate the idea with dimensional analysis: Consider a dimensionless quantity, such as the effective action $\G$ of perturbative quantum gravity, and expand $\G$ in terms of the coupling constant $\ell_P^2$. Every term in the expansion must be dimensionless, and the leading term in the effective action $\G$ is of order $O(1/\ell_P^2)$. Therefore
\be
\G=\frac{\a_{-1}}{\ell^2_P E^2}+\a_0+\a_{\rm log}\log\lt(\ell_P^2 E^2\rt)+\a_1\ell_P^2 E^2+\a_2\ell_P^4 E^4+\cdots.\label{dimanal0}
\ee
The term of type $\ell_P^{2n} E^{2n}\log\lt(\ell_P^2 E^2\rt)$ is also allowed in the expansion. The coefficients $\a_n$ are all dimensionless, while $E$ is an energy scale and it is identified to be the UV cutoff. In order that each term is dimensionless, $\ell_P^{2n}$ has to combine $E^{2n}$. When we look at the UV behavior of the theory by sending $E\to\infty$, there are infinitely many terms in $\Gamma$ that are power-law divergent, and the degree of divergence grows when expanding to higher order. This issue of divergence is a simple consequence of the fact that the coupling constant $\ell_P^2$ is of the dimension $[\mathrm{length}]^2$. This dimensional analysis of \eqref{dimanal0} illustrates that the naive perturbative quantum gravity behaves badly at high-energy. This bad behavior closely relates to the non-renormalizability of quantum gravity. 

Supposing, instead, that $\G$ is computed in LQG on any lattice. The expansion \eqref{dimanal0} can be obtained by a computation similar to the lattice perturbation theory, where the UV cutoff $E$ relates to the lattice spacing $l$ by\footnote{Assuming the spacetime manifold is a 4-torus with Euclidean signature.} 
\be
E=\frac{2\pi}{l}.
\ee
The LQG area spectrum indicates the relation between $l$ and SU(2) spin $j$ by 
\be
l^2= 8\pi \b \ell_P^2j_0,\qquad j_0=\sqrt{j(j+1)}\;,
\ee
where $\b$ is the Barbero-Immirzi parameter. Inserting these ingredients in \eqref{dimanal0}, $\ell_P^2$ cancels in each terms of $\G$
\be
\ell_P^2 E^2=\frac{\pi}{2\b j_0}.
\ee
Then $\G$ is finite order-by-order as far as $j_0\neq 0$
\be
\G =\b j_0\,\a_{-1}+ \a_0+\a_{\rm log}\log(j_0)+\a_1\lt(\frac{1}{\b j_0}\rt)+{\a_2}\lt(\frac{1}{\b j_0}\rt)^2+\cdots .\label{dimanal1}
\ee
where the coefficients $\a_n$ have been re-defined. The power-law and logarithmic divergent terms in \eqref{dimanal0} are translated to $O(1/j_0^n)$ and $\log(j_0)$ in \eqref{dimanal1}. $\G$ is UV finite order by order when $j_0$ is nonzero. In Section \ref{1-loop effective action in long wavelength approximation} and \ref{Beyond the long wavelength approximation}, we will compute explicitly the quantum effective action from the canonical LQG and demonstrate a similar finiteness. We emphasize that, since we are working on a lattice, we don't have a well defined realization UV completeness, as in the continuous dynamics. We also remark that the $1/j$-expansion as in \eqref{dimanal1} is well-known in the literature of spinfoam asymptotics (see e.g. \cite{semiclassical,HZ,Han:2020fil}).

The expansion \eqref{dimanal1} is divergence-free order by order is due to the discreteness of the lattice. As a heuristic argument toward the fundamental discreteness, the LQG area spectrum has a minimal area gap with $j_0=\sqrt{3}/2$, which is a physical UV cutoff, so \eqref{dimanal1} has to be divergence-free. However, our investigation of the quantum effective action in Section \ref{1-loop effective action in long wavelength approximation} gives a more interesting argument: The spacetime geometry $g_{\mu\nu}$ is dynamical in quantum gravity complying to principle of background independence. Thus the quantum area $j$ is dynamical since it quantizes a part of $g_{\mu\nu}$. This is very different from a UV cutoff that is put in by hand as an external parameter. As evident from our analysis, $j_0$ in \eqref{dimanal1} is not a coupling constant or a cutoff but an expectation value of the area operator with respect to certain quantum state, so the value of $j_0$ is determined by the quantum dynamics of the state.\footnote{See recent developments in \cite{Dittrich:2022yoo,Borissova:2022clg} for a different approach constructing the area metric effective action from effective spinfoams.} In other words, how discrete is the spacetime is not put in by hand, rather it is determined by the quantum dynamics of the theory. A convenient way to determine the expectation value $j_0$ is to employ the quantum effective action $\G$ and the quantum equation of motion from $\G$'s variation. As far as the solution $j_0$ is nonzero, which is indeed the case in our analysis, the expansion \eqref{dimanal1} is finite order by order.

\bigskip

This paper is structured as follows. In Section \ref{sec:2}, as a warm up, we introduce the philosophy behind the effective approach, as it it applied in QFT and we review the application of such perturbative QFT methods to Einstein gravity.  In Section \ref{sec:3} we go beyond the standard asymptotic states prescription and construct the coherent state path integral and we introduce the formulation within LGG in the reduced phase space framework. We highlight the differences between the standard QFT and the LQG formulation. Section \ref{sec:4} serves as the setup for the calculation of effective action from the coherent state path integral in LQG around a flat background: we construct the boundary states, the perturbations and the Hessian matrix. In Section \ref{sec:5} we give the expression of the propagator, comparing the classical limit with the numerical evaluation of the short wavelength limit.  Section \ref{1-loop effective action in long wavelength approximation} contains the analytical computation of the long wavelength effective action, while Section \ref{Beyond the long wavelength approximation} is devoted to the numerical investigation of scaling and converge properties beyond the wavelength approximation. These last three sections represent the main results of this work: the Hessian matrix, the propagator and the effective action. Finally, in Section \ref{sec:8} we conclude and give an outlook.  Appendix \ref{Scaling of f(N)} contains details of the computation for the analytical computation of the effective action.

\section{The effective action in perturbation theory}\label{sec:2}
To study renormalization flows and bridge the UV and IR regimes, the effective approach is commonly employed.

The effective approached is based on the separation of   low energy quantum effects from   high energy contributions. Quantum corrections can be then effectively encoded in the low-energy regime.
As a result, effective descriptions are only valid in a specific regime, in particular for energies less than a given scale. More broadly, an \textit{effective action} is one that yields the same results as a given action but involves different degrees of freedom.
Differences with the bare action include that often the effective action has fewer fields, is potentially non-renormalizable, and only has a limited range of validity. When a field is in the full theory but not in the effective action, we say it has been \textit{integrated out}. The advantage of using the effective action over the full theory actions is that by focusing only on the relevant degrees of freedom for a given problem and calculations are often more accessible. 

In this section, we will review the procedure to calculate the effective action in a perturbative way: the computation of functional determinants from the path integral and the use of asymptotic states \cite{Weinberg:1995mt}.

\subsection{Effective actions from functional determinants}
Here we want to integrate over some fields by performing the path integral. The evaluation of the path integral results in the famous $\text{Tr} \log$ (or equivalently the $\log \det$) formula for the Hessian.

In order to compute the effects of quantum corrections in the effective action, we first have to obtain the mean field or saddle point solution about which we will expand. We will start by  the generating functional for a quantum field $\hat \phi$ in $d$ spacetime dimensions
in the presence of a linearly coupled source $J(x)$ given by %
\begin{equation}
    Z[J] =e^{i W[J]}= \int \mathcal{D}\hat\phi\, e^{ S\left[\hat\phi\right]+ i \int \text{d}^dx J(x)\hat\phi(x)}\;.
\end{equation} 
The expectation value of $\hat{\phi}$ in presence of the source $J$ is given by
$$
\frac{\delta W[J]}{\delta J(x)}= \phi(x) \;.\label{WJphi}
$$
The quantum effective action is defined as the Legendre transform of $W[J]$ with $J(x)$ and $\phi(x)$ conjugated Legendre variables:
\begin{equation}\label{eq:eff}
\Gamma[\phi]\equiv W[J_\phi] - \int \text{d}^dx J_\phi(x)\phi(x)\;.
\end{equation} 
The right-hand side of \eqref{eq:eff} is evaluated at the solution $J_\phi(x)$ of \eqref{WJphi}, so $\Gamma$ is a functional of the expectation value $\phi(x)$. The variation of $\G$ gives
\be 
\frac{\delta \Gamma[\phi]}{\delta\phi(x)}=-J_\phi(x)+\int\rmd^d x'\lt[\frac{\delta W[J_\phi]}{\delta J_\phi(x')}\frac{\delta J_\phi(x')}{\delta\phi(x)}-\phi(x')\frac{\delta J_\phi(x')}{\delta \phi(x)}\rt]=-J_\phi(x).
\ee
When the source is turned off $J=0$, we obtain the quantum equation of motion 
\be
\frac{\delta \Gamma[\phi]}{\delta\phi(x)}\Bigg|_{\phi_{\rm sol}(x)}=0\;.
\ee
It implies that the solution $\phi(x)=\phi_{\rm sol}(x)$ to this equation is the (quantum) vacuum expectation value of $\hat{\phi}$.

As a warm-up example of computing $\G[\phi]$ with the background field method, let us work in flat Minkowski spacetime and consider an action\footnote{In our conventions, the imaginary factor $i$ is in the action and not in the exponential in \eqref{eq:Z}.} for a scalar field consisting of a kinetic term and a potential $V(\phi)$
\begin{equation}
    S[\phi]=i \int \text{d}^dx\; \left\{\frac{1}{2}\partial_\mu \phi \partial^\mu \phi -V(\phi)\right\}\;.
\end{equation}
We set $\bar{\phi}(x)$ to be an arbitrary background field and expand $\phi(x)$ as
\begin{equation}
    \phi(x) = \bar \phi(x) + \delta \phi(x)\;,
\end{equation}
it is straightforward to verify that the linear terms in $\delta \phi$ will be cancelled once the saddle point condition is imposed to $\bar{\phi}$. Then, for small fluctuations $\delta \phi$ we can write the generating functional as
\begin{equation}\label{eq:Z}
Z[J] = e^{\frac{1}{\hbar} S\left[\bar \phi\right]+ \frac{i}{\hbar} \int \text{d}^dx J \bar \phi} \int \mathcal{D}(\delta\phi) e^{\frac{i}{\hbar} \int \text{d}^d x\left(\frac{1}{2} \partial^\mu (\delta \phi)\partial_\mu (\delta \phi)-\frac{1}{2} (\delta \phi)^2 V''\left(\bar \phi\right)\right)}\;.
\end{equation}
We have recovered $\hbar$ in the path integral, since the quantum effective action will be expressed as a power series in $\hbar$. The functional integral is Gaussian for small fluctuations, resulting in the following stationary phase approximation in $\hbar$
\begin{equation}
Z[J] = e^{\frac{1}{\hbar} S\left[\bar \phi\right]+ \frac{i}{\hbar} \int \text{d}^dx J(x) \bar \phi} \text{det} (H)^{-1/2}\lt[1+O(\hbar)\rt]\;,
\end{equation}
where we defined the Hessian $H = \partial^2 + V''\left(\bar \phi\right)$. One finds
\begin{equation}
iW[J] = \frac{1}{\hbar}S\left[\bar \phi\right] + \frac{i}{\hbar} \int \text{d}^dx J(x) \bar \phi(x)- \frac{1}{2}\log\det (H)+O(\hbar)\;.
\end{equation}
Importantly, the last term encodes the one-loop quantum corrections coming from integrating out the fluctuations $\delta \phi$. Plugging this in into the effective action one finds
\begin{equation}\label{eq:EA}
i\Gamma\left[\bar \phi\right] = \frac{1}{\hbar} S\left[\bar \phi\right]- \frac{1}{2}\log\det (H) +O(\hbar)\ ,    
\end{equation}
which tells us that the effective action is an expansion around the classical action that includes the quantum fluctuations. 

In order to compute it explicitly, we need to make some assumptions that will allow us to treat the trace. If we assume that the vacuum solution $\bar \phi$ is a constant (local potential approximation), then $V''(\bar \phi)$ is also a constant. Hence, by using the relation $\log \det(H)=\mathrm{Tr}\log(H)$
\begin{equation}
    \text{Tr ln} \left(\partial^2 + V''\left[\bar\phi\right]\right) =  \int\rmd^dx\int \frac{\text{d}^d p }{(2\pi)^d}\;\left(-p^2+ V''\left(\bar\phi\right)\right)
\end{equation}
In order to evaluate the quantum corrections to the equation of motion one defines the effective potential as
\begin{equation}
  V_\text{eff} =V\left(\bar\phi\right)-\frac{i}{2}  \int \frac{\text{d}^d p }{(2\pi)^d}\; \log\left(\frac{-p^2+V''\left(\bar\phi\right)}{-p^2}\right)
\end{equation}
Problematically, this integral has  two divergences: a quadratic and a logarithmic divergence. In order to tame these divergences one can add two counter-terms. Performing a Wick rotation one finds 
\begin{equation}
  V_\text{eff} =V\left(\bar\phi\right)+ \frac{\pi^{d/2}}{\G(d/2)}\int_0^{\Lambda_{UV} }\frac{\text{d} p_E }{(2\pi)^d} p_E^{d-1}\; \log\left(1+\frac{V''(\phi)}{p_E^2}\right) + \text{ counter-terms}
\end{equation}
Expanding the logarithm and keeping only UV-divergent terms, the lowest energy scale is identified with $V''(\bar \phi)$, so one obtains 
\begin{equation}\label{Veffphi111}
  V_\text{eff} =V\left(\bar\phi\right)+ \frac{\Lambda_{UV}^2}{ 32\pi^2}V''(\bar \phi) -\frac{1}{64\pi^2}(V''(\bar \phi))^2\log \left(\frac{\Lambda_{UV}^2}{V''(\bar \phi)}\right) +\text{ counter-terms}
\end{equation}
We highlight at this stage that to fix the counter-terms one needs to impose renormalization conditions. 

Crucially, the investigation of the effective potential and its minima can play a role in the phenomenon of spontaneous symmetry breaking: when computing the effective solution,  due to the quantum corrections the effective potential has a different ground state, which does not share the same symmetries of the original action:
\begin{equation}\label{eq:1lEOM}
    \frac{\delta V_\text{eff}}{\delta \phi} \Bigg|_{\bar \phi_\text{eff}}= \frac{\delta V}{\delta \phi} + \text{quantum corrections} = 0\;.
\end{equation}
As a standard procedure,  in the  one-loop computations the tree-level equations of motion are used. However, in this work we will self-consistently study the one-loop improved equations of motion \eqref{eq:1lEOM}, deriving from the one-loop effective action (see \cite{Branchina:2003kf,Becker:2021pwo,Ferrero:2024yvw} for similar analyses in a QFT context). 

\bigskip

Building on this, it has been questioned whether gravity operates as a well-defined quantum field theory at ordinary energy scales \cite{Donoghue:1994su,Donoghue:1994dn}. 
For this purpose, computations of the one-loop effective action (and its divergences and renormalizability properties) have also been performed for Einstein gravity \cite{tHooft:1974toh,Barvinsky:1983vpp,Parker:2009uva,Becker:2021pwo, Falls:2024noj}.

\subsection{One-loop divergences in perturbative quantum gravity}

It has long been known since the pioneering work of ’t Hooft and Veltman, that pure gravity with a vanishing cosmological constant is a renormalizable theory at one-loop in $d=4$ \cite{tHooft:1974toh}. It is free of logarithmic divergences, while other power law divergences are either not ``seen'' in the minimal subtraction scheme within dimensional regularization, or can be cancelled by going on-shell within momentum regularization or proper time regularization.
 Subsequently, it was observed that one-loop quantum calculations in six dimensions share some features with two-loop calculations in four dimensions. It was found that quantum gravity in six dimensions contains a non-vanishing logarithmic divergence, suggesting that divergences could emerge at two-loops in four dimensions as well. Eventually, Goroff and Sagnotti explicitly calculated the two-loop divergences in four dimensions \cite{Goroff:1985sz,Goroff:1985th}. They demonstrated that pure quantum gravity is a non-renormalizable theory at two-loops. This result was later checked and confirmed by van de Ven \cite{vandeVen:1991gw}. On the other hand, the inclusion of a cosmological constant gives rise to a one-loop logarithmic divergence already in four dimensions, as found by Christensen and Duff \cite{Christensen:1979iy}. This additional divergence, however, can be reabsorbed into the renormalization of the cosmological constant at one-loop.

Here, we will report the well-known result of the computation of the one-loop effective action by means of the proper time (or heat kernel) method with a proper time cutoff $\Lambda_\text{UV}$. This allows to exploit the background field method and to preserve background independence ``à la DeWitt" \cite{DeWitt:1964mxt,DeWitt:1975ys,DeWitt:2003pm}. In particular, this allows as to make a split between a background and a fluctuation:
\begin{equation}
    g_{\mu \nu} = \bar g_{\mu \nu} + h_{\mu\nu}\;.
\end{equation}
Using a standard harmonic (Feynman) gauge fixing and considering both  graviton and  ghosts contributions, one finds  quartic divergences
\begin{equation}
    - \frac{1}{2}\frac{1}{(4\pi)^2}\Lambda_\text{UV}^4 \int \text{d}^4x \sqrt{\bar g}\;,
\end{equation}
 quadratic divergences
\begin{equation}
    - \frac{1}{(4\pi)^2}\Lambda_\text{UV}^2 \int \text{d}^4x \sqrt{\bar g}\left(-\frac{23}{6}\bar R + 10 \Lambda\right)\;,
\end{equation}
and logarithmic divergences
\begin{equation}
    - \frac{1}{(4\pi)^2}\log\left(\frac{\Lambda_\text{UV}^2}{\mu^2}\right) \int \text{d}^4x \sqrt{\bar g}\left(\frac{7}{40}\bar C^2 +\frac{1}{8}\bar R^2+\frac{149}{360}\bar E -\frac{13}{3}\bar R \Lambda + 10 \Lambda^2\right)\;.
\end{equation}
Here we have introduced a scale $\mu$ in order to account for the dimensionful argument of the logarithm, and we have denoted $\bar C$ the  Weyl tensor and $\bar E$ the Euler term.
As renormalization conditions, the power law divergences can be incorporated into the renormalization of the cosmological constant and Newton's constant. The most straightforward approach is to define the renormalized Newton's constant and cosmological constant as follows:
\begin{equation}
\frac{1}{G_R} = \frac{1}{G}-\frac{1}{\pi} \left(\frac{23}{6}\Lambda_\text{UV}^2 + \frac{13}{3} \Lambda\log\left(\frac{\Lambda_\text{UV}^2}{\mu^2}\right) \right)\;,
\end{equation}
\begin{equation}
\frac{\Lambda_R}{G_R} = \frac{\Lambda}{G}-\frac{1}{4\pi} \left(\Lambda_\text{UV}^4+20 \Lambda \Lambda_\text{UV}^2 + 20\Lambda^2\log\left(\frac{\Lambda_\text{UV}^2}{\mu^2}\right) \right)\;.
\end{equation}
The bare couplings $G$ and $\Lambda$ must be tuned  so that the renormalized $G_R$ and $\Lambda_R$ align with observed values. In particular, to match observations \cite{SupernovaSearchTeam:1998fmf}, $\Lambda$ must be set so that  $\Lambda_R$ is either zero or, at the very least, significantly smaller than the renormalized Planck mass,   $(8\pi G_R)^{-1/2}$. This problem is known as the  fine-tuning problem of the cosmological constant \cite{Weinberg:1988cp}.

The remaining logarithmic divergences involve terms not already present in the bare action. At first glance, this might suggest that the theory is non-renormalizable. However, this is not necessarily the case. In fact, physical results are obtained by going on-shell, which involves expanding the metric around a saddle point of the action. Concretely, this requires the background metric to satisfy the equations of motion. In particular, the outcome depends on whether the renormalized cosmological constant is zero or nonzero.
First, consider pure gravity with a zero cosmological constant. The equations of motion enforce that the Ricci tensor must vanish, rendering the potentially divergent terms identically zero. Consequently, the theory is one-loop renormalizable. Additionally, without a cosmological constant, the field equations for pure gravity ensure that the Ricci tensor vanishes on-shell. Therefore, working in the Riemann basis, the only term quadratic in the curvature that does not vanish on-shell is the $R^{\mu \nu \rho \sigma}R_{\mu \nu \rho \sigma}$. However, in four dimensions, this term can be locally expressed as a total derivative (i.e., the Euler topological term) along with additional terms that are quadratic in the Ricci tensor and Ricci scalar, which vanish on-shell.

Hence, in the case of pure gravity, the logarithmic divergences vanish on-shell, so there must be a field redefinition that removes those terms. For instance one could choose
\begin{equation}
\delta g_{\mu \nu} \to\frac{1}{(4\pi)^2}\log\left(\frac{\Lambda_\text{UV}^2}{\mu^2}\right) \frac{1}{20}\left(7 \bar R_{\mu \nu} -\frac{11}{3}\bar g_{\mu \nu} \bar R\right)\;,
\end{equation}
which would eliminate the logarithmic divergences.

The situation is different when matter is present. The equation of motion implies that the coefficient of the logarithm is proportional to the square of the energy-momentum tensor, and such terms are not present in the bare Lagrangian. In general, there will be genuinely divergent terms that do not match the form of the operators originally present in the action. As a result, gravity coupled to matter is  non-renormalizable at one loop.

If the cosmological constant is not zero, the equation of motion implies $\bar R = 4 \Lambda$. As a consequence there will be additional terms contributing to Ricci tensor squared and Ricci scalar squared which are nonzero and will contribute to the logarithmic renormalization of the cosmological constant \cite{Christensen:1979iy}.  The rest of the argument is unchanged, so pure gravity with cosmological constant is also one-loop renormalizable.

To conclude this section, let us emphasize once more that off-shell quantities are in general renormalization scheme- as well as gauge-dependent and that such dependencies cannot enter in physical observables.  This is another way of saying that the logarithmic divergence of pure gravity at one-loop is unphysical. In particular, recently, it has been shown how adopting a non-minimal scheme in order to account for the on-shellness of the fields, the divergences present for Einstein gravity at one-loop can be reabsorbed and that the renormalized couplings do not depend on unphysical choices \cite{Falls:2024noj,Kluth:2024lar}.

\section{Coherent state path integral} \label{sec:3}

In the previous section we have reviewed the status of the divergences arising from the application of QFT methods to Einstein gravity on an arbitrary background. Specifically, we have examined the one-loop divergences and reviewed the methods used to tame those divergences. As a consequence of  the failure of those methods beyond one-loop, a way around the lack of perturbative renormalizability has been developed in the past decades. The main idea is to study whether there are chances that a quantum theory of gravity might instead be non-perturbatively renormalizable (as in Asymptotic Safety) or in general perturbatively renormalizable in a general background within a particular renormalization scheme via the background field method in a background independent way \cite{Weinberg:1980gg,Ferrero:2024rvi}.

In this work we will explore  a new direction: we inherit the non-perturbativity not at the level of the functional integral, but from the non-perturbative quantization within LQG. With this accomplished, the investigation of the divergences is achieved via standard QFT perturbative methods.
Concretely, our purpose is to compute the one-loop effective action from the coherent state path integral derived from a non-perturbative quantization of gravity achieved from LQG. 

The purpose of making use of perturbative methods  is twofold.  First, many of the “problems” encountered in non-perturbative RG approaches can be discussed and resolved relatively easily within the framework of perturbation theory, where it can be straightforwardly tracked how  gauge and parametrization dependencies propagate. In fact,  in non-perturbative methods unwanted parametric dependencies combine in the renormalization of the source term for the operator corresponding to the equations of motion, which, in turn, implies that the natural argument of the effective action is not simply the sum of the background metric and the expectation value of the fluctuation, but, rather, a more complicate functional operator. Second, we wish to establish a setup which can be generalized to higher orders, analogously as it is done in other fields of applications of QFT methods.

\subsection{Effective action and coherent states: use in QFT and in LQG}\label{Effective action and coherent states}
We have seen that the  effective action is a useful tool extensively employed in various fields, particularly in particle physics. It has two key properties: (i) its stationary solution defines the vacuum state, and (ii) it serves as the generating functional for the complete set of off-shell one-particle-irreducible (1PI) Green's functions. Investigations on the on-shell properties of the effective action extract all the observable information out
of the effective action. Ultimately, for the quantum system, the role of the classical action is played by the effective action. 

In the typical functional integral formulation for the infinite time interval $t_i = -\infty$ and $t_f = +\infty$,
the boundary states are naturally taken as the vacuum by
the use of the asymptotic states prescription (we refer the reader to \cite{Weinberg:1995mt} for the construction of the S-matrix based on in- and out-states and to Chapter 23 in \cite{DeWitt:2003pm} for a general treatment). In this subsection, we aim to generalize this prescription by means of non-trivial boundary states.
In particular, the path integral approach we have employed so far is a powerful tool but is not closely aligned with theories based on a canonical quantization.  Here, we introduce a more general framework based on the concept of coherent states \cite{Zhang:1999is}.

Let us now study the on-shell expansion of the effective action for a generic field \eqref{eq:eff}.
The solution of the stationary condition
\begin{equation}\label{eq:stationary}
\frac{\delta \Gamma[\phi]}{\delta \phi (x)}\Big|_{\phi = \phi_0} = 0\
\end{equation}
determines the vacuum expectation value of the field. We
notice, however, this statement is based on the fact that
certain boundary conditions have been assumed on the
functional integral in \eqref{eq:eff}. Hence, by clarifying the boundary contributions, one should write
\begin{equation}\label{eq:00}
    \text{exp}(i W_{00}[J])\equiv \langle 0 \Big | T \text{exp} \left(i \int_{-\infty}^\infty \text{d}^dxJ(x)\hat \phi (x)\right)\Big| 0\rangle\;.
\end{equation}
From this, the stationary solution is obtained via
\begin{equation}
    \phi_0 =\langle0|\hat \phi (x)|0\rangle\;.
\end{equation}
One could ask then whether there exist another solution to \eqref{eq:stationary} in the form $\phi (x) =\phi_0 (x) + \Delta \phi (x)$, where $\Delta\phi (x)$ is an excitation over the vacuum solution ($\phi (x)$ is still a solution of \eqref{eq:stationary} but does not necessary satisfy the same boundary condition as the excitation $\Delta \phi (x)$, see \cite{Berezhiani:2023uwt} for a related discussion on squeezed coherent states). 
For the sake of clarity, we start choosing a zero background solution $\phi_0 = 0$.  Then it can be shown that the variation $\Delta \phi (x)$ from the vacuum solution can be reduced to the changes of the initial and final
states on the right-hand side of \eqref{eq:00} and, by this observation, we clarify the relation between the effective
action and the connected S-matrix element.
First of all  we want to clarify the physical meaning of $\Delta \phi (x)$. The total form of $\Delta \phi (x)$ can be expressed in terms of the coherent states $\theta^\pm$ \cite{Fukuda:1988ka,Komachiya:1990mq} as 
\begin{equation}
    \Delta \phi (x) = \langle \theta^- | \hat \phi (x)| \theta^+ \rangle  + \text{ disconnected terms}\;.
\end{equation}
Now, we introduce the generating functional
\begin{equation}\label{eq:Wtheta}
  \text{exp} \left(i W_{\theta^- \theta^+}[J]\right)  \equiv \langle \theta^- \Big | T \text{exp} \left(i \int_{-\infty}^\infty \text{d}^dxJ(x)\hat \phi (x)\right)\Big| \theta^+\rangle\;,
\end{equation}
which corresponding to a path integral with boundary states $\theta^\pm$ instead of the vacuum. The boundary states satisfy
\begin{equation}
\frac{\delta W_{\theta^- \theta^+}}{\delta J(x)} \bigg|_{J = 0} =  \langle \theta^- | \hat \phi (x)| \theta^+ \rangle \equiv \phi_*\;. 
\end{equation}
From this expression we find that
looking for another solution of \eqref{eq:stationary} around the vacuum
solution can be interpreted as changing the initial and
final states into the form of the coherent states.

Finally, one can also define the following effective action by using \eqref{eq:Wtheta}:
\begin{equation}
\Gamma_{\theta^- \theta^+}[\phi_*] \equiv  W_{\theta^- \theta^+}[J]- \int \text{d}^dx J(x) \phi_*(x)\;.
\end{equation}
$\Gamma_{\theta^- \theta^+}[\phi_*]$ is the quantum effective action based on the path integral with boundary coherent state $\theta^\pm$. The effective field $\phi_*$ satisfies the boundary condition defined by $\theta^\pm$, different from the asymptotic vacuum boundary condition. 
The expansion $\Gamma_{\theta^- \theta^+}[\Delta \phi]$
is then  the on-shell expansion of the
effective action in the sense that all the terms  are
projected onto the mass shell. Using the LSZ (Lehmann–Symanzik–Zimmermann) reduction and considering a one-particle excitation $\Delta\phi^{(1)}$ formula one gets
\begin{equation}
\begin{aligned}
   \Gamma_{\theta^- \theta^+}[\Delta \phi]&=W_{\theta^- \theta^+}[J = 0]=-i \langle\theta^-|\theta^+\rangle _{(\text{connected})}\\&=\Gamma_{00}[\phi_0 =0]-i \langle1^-|1^+ \rangle + \sum_{n = 3}^\infty \frac{1}{n!}\left( \tilde W_{00}^{(n)}\right)_{x_1,\cdots, x_n} \Delta\phi^{(1)}(x_1) \cdots \Delta \phi^{(1)} (x_n)\;.
   \end{aligned}
\end{equation}
Here the tilde demotes the $W_{00}^{(n)}$ with amputated external legs and
\begin{equation}
    |1^{+(-)}\rangle \equiv \int \text{d}^3 k \;a^\dagger_{\text{in(out)}} (\vec {k})|0\rangle\; ,\qquad \text{where } a^\dagger_\text{in} \equiv C^{+}(\vec {k}) \hat a_\text{in}^\dagger\;,a_\text{out} \equiv C^{-}(\vec {k}) \hat a_\text{out}
\end{equation}
where $\hat a_{\text{in(out)}}$ are the creation and annihilation operators for  $\hat \phi_{\text{in(out)}}$ and   $C^{+(-)}$ the Fourier coefficients for the one-particle excitation $\Delta  \phi^{(1)}$
\begin{equation}
\Delta \phi^{(1)} (x) = \int \text{d}^3 k \;\left(C^+(\vec{k})e^{-ik\cdot x}+C^-(\vec{k})e^{ik\cdot x}\right)\;.
\end{equation}
Thus, we can conclude that $  \Gamma_{\theta^- \theta^+}[\Delta \phi]$ corresponds to the generating functional of the connected S-matrix elements.

The relation between $\Gamma_{\theta^- \theta^+}$ and $\Gamma$ can be made clear by exploiting a formulation in terms
of a source at the boundary. Introducing  an additional source at the initial  and the final time 
which plays the role of changing the vacuum state into
the desired coherently excited state. The effective action
is calculated under the presence of this type of source
while the boundary state is the vacuum. In this way one can  define a new effective action whose boundary states is the vacuum state instead of the coherent states \cite{DeWitt:2003pm}. The source term  would be
\begin{equation}
    \int \text{d}^4x K(x)\hat \phi (x)\;, \text{ where} \qquad K(x) = - \int_{-t_1}^{t_2} \text{d}^4x' \Delta \phi (x') \frac{\delta{\Gamma_{00}}}{\delta \phi(x')\delta \phi(x)}
\end{equation}
Now we have the following pair of equations
\begin{equation}
\frac{\delta{\Gamma_{00}}}{\delta \phi(x)}\Bigg|_{\phi = \phi_0} = 0\; , \qquad \frac{\delta{\Gamma_{00}}}{\delta \phi(x)}\Bigg|_{\phi = \phi_0+ \Delta \phi} = -K(x)
\end{equation}
Hence, for $t \in [-t_1, t_2]$ we obtain the $\Delta\phi$ solution, while, for $t<-t_1$ or $t>t_2$ the solution becomes $\phi_0$, assuming that at $t = \pm \infty$ the state is the vacuum state. Finally one  takes the limit $t_{1,2}\to \infty$ at the end. Hence, the introduction of such a source $K(x)$ is equivalent to setting the boundary condition which leads
to the coherent states.  After taking
the limit $t_{1,2}\to \infty$, the effect of the  introduced
source term $K(x)$ becomes to excite the vacuum state at
the boundary $t\to \pm \infty$. Finally,  the new effective action $\Gamma_{00}[\phi, K]$ can be related to the previously introduced one as
\begin{equation}
    \Gamma_{00}[\phi, K] = \Gamma_{00}[\phi] +   \int \text{d}^4x K(x) \phi (x)\;.
\end{equation}
The derivation of the explicit relationship between $\Gamma_{00}[\phi, K]$ and $\G_{\theta^-\theta^+}$ is more involved. Here we will sketch the main steps. By using the definition of $\Gamma_{00}[\phi,K]$ with $J = 0$ and using again the LSZ reduction formula one finds
\begin{equation}
    \text{exp}\left(i \Gamma_{00}[\Delta \phi, K]\right) =  \text{exp}\left(-\int \text{d}^3k\;2k^0\;(2\pi)^3\; C^{-}(\vec{k})C^{+}(\vec{k})\right)\text{exp}\left(i \Gamma_{\theta^- \theta^+}[\Delta \phi]\right)
\end{equation}
Now, considering the case $\phi_0 \neq 0$, the generating functional $W_{\theta^- \theta^+}$ is defined with the operator $\hat \phi -\phi_0$.
Using the relation
\begin{equation}
\begin{aligned}
    &\text{exp}\left(i W_{\theta^-\theta^+}[J]\right) \equiv \bigg\langle \theta^-\bigg| T \;\text{exp}\left(i\int\text{d}^4x\;J(x)\big(\hat \phi(x)-\phi_0(x)\big)\right)\bigg|\theta^+\bigg\rangle\\&\qquad=\text{exp}\left(\int \text{d}^3k\;2k^0\;(2\pi)^3\; C^{-}(\vec{k})C^{+}(\vec{k})\right) \bigg\langle 0\bigg| T \;\text{exp}\left(i\int \text{d}^4x\;(J(x)+K(x))\big(\hat \phi(x)-\phi_0(x)\big)\right)\bigg|0\bigg\rangle
    \end{aligned}
\end{equation}
we get the relation
\begin{equation}
    \begin{aligned}
    \Gamma_{\theta^-\theta^+}[\phi^*] &\equiv     W_{\theta^-\theta^+}[J] -\int \text{d}^4 x J(x)\phi^* (x)\\& = W_{00}[J+K]- i \int \text{d}^3k\;2k^0\;(2\pi)^3\; C^{-}(\vec{k})C^{+}(\vec{k})  -\int \text{d}^4 x J(x)\phi^* (x)
        \end{aligned}
\end{equation}
Using the Legendre transformation of $W_{00}[J+K]$ we get the desired relation
\begin{equation}
    \begin{aligned}
    \Gamma_{\theta^-\theta^+}[\phi^*] = \Gamma_{00}[\phi^*] -i\int \text{d}^3k\;2k^0\;(2\pi)^3\; C^{-}(\vec{k})C^{+}(\vec{k}) + \int \text{d}^4x\; K(x) \phi^*(x)
        \end{aligned}
\end{equation}
Setting $J=0$, then this becomes the generating functional of the connected S-matrix elements, which is constructed directly from the usual effective action $\Gamma_{00}[\phi]$.

\bigskip

Starting from the next subsection, we introduce an LQG path integral analog of \eqref{eq:Wtheta}, where $\theta^\pm$ are (Hamiltonian evolution of) LQG gauge invariant coherent states. Following a similar way as in \eqref{eq:EA}, the LQG analog of $\G_{\theta^- \theta^+}[\phi_*]$ will be computed in Section \ref{1-loop effective action in long wavelength approximation} from the path integral formula with boundary coherent states. 

The effective field $\phi_*(x)=\langle\theta^- |\hat{\phi}(x)|\theta^+\rangle$ with vanishing external source satisfies the quantum equation of motion
\be
\frac{\delta\G_{\theta^-\theta^+}[\phi_*]}{\delta\phi_*(x)}=0. \label{Qeomtheta}
\ee
Suppose we start from \eqref{eq:Wtheta} with a pair of $\theta^\pm$ and obtain the quantum effective action $\G_{\theta^-\theta^+}[\phi_*]$. If the solution to \eqref{Qeomtheta} exists, the solution gives the dynamics of $\langle\theta^- |\hat{\phi}(x)|\theta^+\rangle$. However, for certain boundary states $\theta^\pm$, the solution may not exist. Actually, the existence of solution to \eqref{Qeomtheta} gives the constraint to the boundary states $\theta^\pm$: Since $W_{\theta^-\theta^+}[0]=\Gamma_{\theta^- \theta^+}[\phi_*] $ is a function of $\theta^\pm$, the existence of solution to \eqref{Qeomtheta} picks up $\theta^\pm$ as the critical point of $\exp(iW_{\theta^-\theta^+}[0])$.

At this stage, it is crucial to highlight the fundamental difference in approach compared to standard QFT computations. Unlike the field-theoretic method, we do not begin with a vacuum asymptotic state solution, nor do we aim to establish boundary conditions for excitations of such a state. Instead, our focus is on identifying consistent boundary conditions that satisfy the quantum equations of motion. Whether these two approaches can be reconciled remains an open question and a detailed discussion of the vacuum state in LQG \cite{Ashtekar:1993wf,Ashtekar:1994mh} lies beyond the scope of this paper.

\subsection{Coherent-state path integral from LQG}\label{Coherent state path integral from LQG}

So far, we have not considered physical states generated by the Hamiltonian. This is the purpose of this subsection, where we will specify the setup to the LQG coherent state path integral. It has been shown, how using coherent states, one can derive a new discrete path integral expression for the transition amplitude generated by the physical Hamiltonian $\hat{\bf H}$ for gravity coupled to dust \cite{Giesel:2007wn,Han:2019vpw}.

Consider a finite cubic lattice $\g$ discretizing the spatial slice that is a 3-torus $\mathbb{T}^3$. $E(\g)$ and $V(\g)$ denote the set of edges and vertices in $\g$. The LQG Hilbert space on $\g$ has an over-complete basis consisting of (gauge invariant) complexifier coherent states $\Psi^t_{[g]}$ \cite{Thiemann:2000bw}. The transition amplitude between a pair of $\Psi^t_{[g]}$ and $\Psi^t_{[g']}$ is
\begin{equation}\label{eq:amplitude}
  A_{[g],[g']}=  \langle\Psi^t_{[g]}|\,\exp[-\frac{i}{\hbar}\hat{\bf H} T]\, |\Psi^t_{[g']}\rangle\;,
\end{equation}
where $T$ is the physical time parametrized by the dust field \cite{Giesel:2007wi,Giesel:2007wn}. $g$ is the coherent state label $g=\{g(e)\}_{e\in E(\g)}$ with 
\be
g(e)=e^{-ip^a(e)\t^a/2}h(e)=e^{-ip^a(e)\t^a/2}e^{\theta^a(e)\t^a/2}\in\Slc
\ee
is the holomorphic coordinate on the LQG phase space of holonomies $h(e)$ and gauge covariant fluxes $p^a(e)$, $\t^a=-i(\text{Pauli matrix})^a$, $a=1,2,3$. The gauge invariant coherent states can be written in terms of the normalized non-gauge-invariant coherent states  $\psi^t_g$ by
\be
\Psi^t_{[g]}(h)
&=&\int_{\mathrm{SU(2)}^{|V(\g)|}}\rmd h\prod_{e\in E(\g)}{\psi}^t_{h_{s(e)}^{-1}g(e)h_{t(e)}}\lt(h(e)\rt),\quad \rmd h=\prod_{v\in V(\g)}\rmd\mu_H(h_v).\label{gaugeinv}
\ee
where $\text{d}\mu_H$ is the Haar measure. Moreover,  $t=\ell_P^2/a^2$ is the the dimensionless semiclassicality parameter, where $\ell_P^2=\hbar\kappa$ is the Planck scale, and $a$ is a length unit.
As a discrete path integral formula, it can be written as an integral over $N+1$ intermediate states $g_i\in\Slc^{|E(\g)|}$ and $h=\{h_v\}_{v\in V(\g)}\in \Su^{|V(\g)|}$, in order to ensure SU(2) gauge invariance:
\begin{equation}
{A_{[g],[g']}}=\int\rmd h\prod_{i=1}^{N+1}\rmd g_i\,\nu[g]\, e^{S[{g},h]/t},\label{LQGPI}
\end{equation}
where $\|\psi^t_{g}\|$ is invariant under the gauge transformation of $g$. $\nu[g]$ is a path integral measure
\be
\nu[g]=\prod_{i=0}^{N+1} \prod_{e \in E(\gamma)}\left[\frac{\mathrm{arccosh}\left(x_{i+1,i}(e)\right)}{\sinh \left(\mathrm{arccosh}\left(x_{i+1,i}(e)\right)\right)}\sqrt{\frac{\sinh \left(p_{i+1}(e)\right)}{p_{i+1}(e)} \frac{\sinh \left(p_{i}(e)\right)}{p_{i}(e)}} +O(t^\infty)\right]\;.
\ee
The path integral formula is obtained by applying \eqref{eq:amplitude} with \eqref{gaugeinv} and discretizing time $T= N \Delta \tau$ with large $N$ and infinitesimal $\Delta \tau$. One can read off from the path integral
$S[{g},h]$, the action for LQG:
\be
S[g,h]&=&\sum_{i=0}^{N+1} K\left(g_{i+1}, g_{i}\right)-\frac{i \kappa}{a^{2}} \sum_{i=1}^{N} \Delta \tau\frac{\langle\psi_{g_{i+1}}^{t}\mid \hat{\mathbf{H}}\mid \psi_{g_{i}}^{t}\rangle}{\langle\psi_{g_{i+1}}^{t} \mid \psi_{g_{i}}^{t}\rangle}\\
K\left(g_{i+1}, g_{i}\right)&=&\sum_{e \in E(\gamma)}\left[z_{i+1, i}(e)^{2}-\frac{1}{2} p_{i+1}(e)^{2}-\frac{1}{2} p_{i}(e)^{2}\right]\;,
\ee
Here 
\begin{equation}
z_{i+1,i}(e)= \text{arccosh}\left(x_{i+1,i}(e)\right),\quad x_{i+1,i}(e)=\frac{1}{2}\text{tr}\left[g_{i+1}(e)^\dagger g_{i}(e)\right].
 \end{equation}
The discrete path integral remains well-defined as a finite integral as long as both $\g$ and $N$ is finite, even if arbitrarily large. In the semiclassical limit 
$t\to 0$, we can look for the saddle point solution. In this approximation, as for the one-loop effective action, the integral is primarily dominated by contributions from the solutions of the equations of motion derived from the variational principle.

Following the standard field theoretical treatment of path integral, when $N\to \infty$  we focus on the evolution that $g_i$ and $g_{i+1}$ are sufficiently close, and we perform the following approximation 
\be 
\frac{\langle\psi_{g_{i+1}}^{t}\mid\hat{\mathbf{H}}\mid \psi_{g_{i}}^{t}\rangle}{\langle\psi_{g_{i+1}}^{t} \mid \psi_{g_{i}}^{t}\rangle}\simeq {\langle\psi_{g_{i}}^{t}\mid \hat{\mathbf{H}}\mid \psi_{g_{i}}^{t}\rangle}=\mathbf{H}[g_i]+O(t),\label{hamiltonExp}
\ee
where the leading order of the expectation value gives the classical physical Hamiltonian $\mathbf{H}[g_i]$ on the lattice. The next-to-leading order in $O(t)$ is studied in \cite{Zhang:2021qul} and will be useful in Section \ref{Quantum equations of motion}. In the following, we adopt this approximation in the path integral formula as the starting point for deriving the one-loop effective action. While this approach lacks mathematical rigor, it is anticipated that the future developments by employing the rigorous definition of the coherent-state Hamiltonian path integral in e.g. \cite{Klauder2011,DeWitt:2003pm}, might address these limitations.

When we couple the dynamical field $p^a(e),\theta^a(e)$ to the external source in the path integral formula \eqref{LQGPI}, we obtain the LQG analog of the generating functional \eqref{eq:Wtheta}. In the following, we perform a perturbative computation of the quantum effective action for LQG.

\section{Background and perturbations}\label{sec:4}

\subsection{Boundary states and Minkowski background}

As a next step, we need to specify the initial and final states consistent with our chosen background symmetries. We choose to work around Minkowski spacetime.
To describe the suitable initial and final states, we orient edges in the cubic lattice $\g$ such that every vertex $v$ connects to 3 outgoing edges $e_{I}(v)$ (oriented toward the $I=1,2,3$ spatial directions) and 3 incoming edges $e_{I}(v-\hat{I})$. We assign the semiclassical holonomy and flux along the edges by
\be
h(e_I(v))= 1,
\quad p^a(e_I(v))=p\,\delta^a_I.\label{flatbd}
\ee
Here $p>0$ is a constant proportional to the constant area associated to every edge. These semiclassical data corresponds to a flat slice in Minkowski spacetime: The constant $p^a(e_I(v))$ implies the flat spatial geometry, and the trivial $h(e_I(v))$ corresponds to the vanishing extrinsic curvature. 

We apply the data \eqref{flatbd} to both the inital and final states $g$ and $g'$ in \eqref{LQGPI}:
\be
g(e_I(v))=g'(e_I(v))=e^{-{i}p\frac{\t^I}{2}}=n_I\,e^{-{i} p\frac{\t^3}{2}}\,n_I^\dagger, \label{nen}
\ee
where $n_{I=1,2}\in\Su$ rotates $(0,0,1)$ to $(1,0,0)$ or $(0,1,0)$ while $n_{I=3}$ is the identity.

The initial and final data \eqref{nen} uniquely determine a solution to equations of motion from $S[g,h]$ \cite{Han:2019vpw}, up to a global sign-flip of $h_v$. The solution is $h_v=\pm1$ and $g_i(e_I(v))$, where $i=0,\cdots,N+2$ labels time steps ($g_0=g'$, $g_{N+2}=g$, and $\pm$ is global on $\g$), are constant and given by \eqref{nen} at all time steps. The solution gives \eqref{flatbd} at all time and implies that the spatial slice $\mathbb{T}^3$ at all time are flat and of vanishing extrinsic curvature. The solution endows a flat geometry to the 4-dimensional hypercubic lattice made by the discrete time evolution of $\g$. The spatial lattice spacing is proportional to $\sqrt{p}$, and the spacing along time direction is $\Delta\t$. If we define the coordinate $\vec{x}$ on $\mathbb{T}^3$ such that the coordinate length of the spatial edge is a constant $\mu$, the flat metric corresponds to the solution is 
\be  
\rmd s^2=-\rmd \t^2+P_0\sum_{I=1}^3(\rmd {x}^I)^2, \qquad p=\frac{2 \mu^{2}}{\beta a^{2}}P_{0}.\label{flatmetric111}
\ee
The relation between $p$ and $P_0$ is obtained from the formula of gauge covariant flux (see e.g \cite{Han:2020iwk}). We use the solution $g_i(e_I(v))=g(e_I(v))$ as the background on which perturbations of holonomies and fluxes are performed in Section \ref{Perturbations}. The measure factor $\nu[g]= 1+O(t^\infty)$ at the background.

Since the path integral formula \eqref{LQGPI} is invariant under the SU(2) gauge transformations of $g$ and $g'$, the initial and final data \eqref{nen} are equivalent to the modifications by gauge transformations 
\be  
g(e)\to u_{s(e)}^{-1}g(e)u_{t(e)},\qquad g'(e)\to w_{s(e)}^{-1}g'(e)w_{t(e)},
\ee
where $u_v,w_v\in\Su$. The solution to the equations of motion is given by $g_i(e)=g(e)$ for all $i=1,\cdots,N+2$ and $h_v=\pm w_vu_v^{-1}$, where $\pm$ is global on $\g$.

\subsection{Perturbations}\label{Perturbations}

Given the background corresponding to Minkowski spacetime, the perturbations $\mathcal{X}^{a}\left(e_{I}(v)\right)$ and $\mathcal{Y}^{a}\left(e_{I}(v)\right)$ of holonomies and fluxes are defined as follows:
\be
\theta^{a}\left(e_{I}(v)\right)= \mathcal{X}^{a}\left(e_{I}(v)\right), && p^{a}\left(e_{I}(v)\right)=\frac{2 \mu^{2}}{\beta a^{2}}\left[ P_{0} \delta_{I}^{a}+\mathcal{Y}^{a}\left(e_{I}(v)\right)\right],\label{perturb1}
\ee
where $p^a$, $P_0$, $\mathcal{X}^{a}$, and $\mathcal{Y}^{a}$ are all dimensionless. We introduce a vector $V^\rho(v)$ as the short-hand notation:
\be
V^\rho(v)=\Big(\mathcal{Y}^{a}\left(e_{I}(v)\right), \mathcal{X}^{a}\left(e_{I}(v)\right) \Big),\quad \rho=1,\cdots 18
\ee
The dictionary between $V^\rho(v)$ and $\cx^a(e_I(v)),\cy^a(e_I(v))$ is given below:
\be
V^1=\cy^1(e_1),\quad &V^2=\cy^2(e_1),&\quad V^3=\cy^3(e_1)\\
V^4=\cy^1(e_2),\quad &V^5=\cy^2(e_2),&\quad V^6=\cy^3(e_2)\nonumber\\
V^7=\cy^1(e_3),\quad &V^8=\cy^2(e_3),&\quad V^9=\cy^3(e_3)\nonumber\\
V^{10}=\cx^1(e_1),\quad &V^{11}=\cx^2(e_2),&\quad V^{12}=\cx^3(e_3)\nonumber\\
V^{13}=\cx^2(e_1),\quad &V^{14}=\cx^3(e_1),&\quad V^{15}=\cx^3(e_2)\nonumber\\
V^{16}=\cx^1(e_2),\quad &V^{17}=\cx^1(e_3),&\quad V^{18}=\cx^2(e_3)\nonumber
\ee
$V^\rho(v)$ captures all perturbative degrees of freedom of holonomies and fluxes.

Because we work with cubic lattice $\g$ with constant coordinate spacing $\mu$ and periodic boundary, it is convenient to make the following lattice Fourier transformation\footnote{Note that here we absorb $\frac{1}{L^3}$ to $\tilde{V}^{\rho}$ in the conventional Fourier transform such that it is length dimension 0.}:
\be
V^\rho(\tau,v)=V^{\rho}(\tau, \vec{n})=\sum_{\vec{k}\in(\frac{2\pi }{L}\mathbb{Z})^3,\ |k^I|\leq \frac{\pi}{\mu}}  e^{i \mu \vec{k} \cdot \vec{n}} \tilde{V}^{\rho}(\tau, \vec{k}), \qquad \vec{n} \in \mathbb{Z}^3.\label{fourierksig}
\ee
where both $\vec{n}$ and $\vec{k}$ have periodicity $n^I\sim n^I+L/\mu$ and $k^I\sim k^I+\frac{2\pi}{\mu}$ ($I=1,2,3$), so the sum $\sum_{\vec{k}}$ has the UV cutoff $|k^I|\leq \frac{\pi}{\mu}$ ($L/\mu$ is assumed to be an interger). Eq.\eqref{fourierksig} can also be expressed as below when we write $\vec{k}=\frac{2\pi }{L}\vec{m}$, and $L=\mu \cn$ where $\cn$ is the total number of vertices along each direction:
\be
V^\rho(\tau,v)=V^{\rho}(\tau, \vec{n})=\sum_{\vec{m}\in\mathbb{Z}(\cn)^3} \prod_{I=1}^3 e^{\frac{2\pi i}{\cn} {m}^I {n}^I} \tilde{V}^{\rho}(\tau, \vec{m}).\label{fouriermn}\\
\tilde{V}^{\rho}(\tau, \vec{m})=\frac{1}{\cn^3}\sum_{\vec{n}\in\mathbb{Z}(\cn)^3}   \prod_{I=1}^3 e^{-\frac{2\pi i}{N} {m}^I {n}^I} {V}^{\rho}(\tau, \vec{n})
\ee
where $\mathbb{Z}(\cn)$ are integers in $[-\cn/2,\cn/2-1]$ if $\cn$ is even, or in $[-(\cn-1)/2,(\cn-1)/2]$ if $\cn$ is odd.

The Hamiltonian $\mathbf{H}[g]$ is expanded to quadratic order in $V^{\rho}$:
\begin{align}
   \mathbf{H}[g] = \sum_{\vec{n}} \mathcal{C}^{(2)} (\vec{n}) + {O}(V^3), \qquad \mathcal{C}^{(2)} (\vec{n})=\sum_{\delta \vec{n}_1} \sum_{\delta \vec{n}_2}  V^{\rho}(\vec{n}+\delta \vec{n}_1) V^{\sigma}(\vec{n}+\delta \vec{n}_2) f_{\rho \sigma}^{\delta \vec{n}_1 , \delta \vec{n}_2}(P_0) .
\end{align}
The appearance of $\delta \vec{n}_{1},\delta \vec{n}_{2}$ is due to the holonomies in $\mathbf{H}[g]$ along several minimal loops, which involves holonomies associated to vertices $\vec{n}+\delta \vec{n}_{1,2}$. By Fourier transformation, we obtain
\begin{align}
    \mathbf{H}[g] &=  \sum_{\vec{n}}\sum_{\delta \vec{n}_1} \sum_{\delta \vec{n}_2}\sum_{\vec{k}} \sum_{\vec{k}'} e^{i \mu \vec{k} \cdot (\vec{n}+\delta \vec{n}_1)} e^{i \mu \vec{k}' \cdot (\vec{n}+\delta \vec{n}_2)} \tilde{V}^{\rho}(\vec{k}) \tilde{V}^{\sigma}(\vec{k}') f_{\rho \sigma}^{\delta \vec{n}_1 , \delta \vec{n}_2}(P_0) \\
    &= \sum_{\vec{k}} \sum_{\vec{k}'} 
 \sum_{\vec{n}}e^{i \mu (\vec{k}+\vec{k}') \cdot \vec{n} }\sum_{\delta \vec{n}_1} \sum_{\delta \vec{n}_2} e^{i \mu \vec{k} \cdot \delta \vec{n}_1} e^{i \mu \vec{k}' \cdot \delta \vec{n}_2} \tilde{V}^{\rho}(\vec{k}) \tilde{V}^{\sigma}(\vec{k}') f_{\rho \sigma}^{\delta \vec{n}_1 , \delta \vec{n}_2}(P_0) \\
 &=  \sum_{\vec{k}} \sum_{\delta \vec{n}_1} \sum_{\delta \vec{n}_2} e^{i \mu \vec{k} \cdot (\delta \vec{n}_1 - \delta \vec{n}_2)} \tilde{V}^{\rho}(\vec{k}) \tilde{V}^{\sigma}(-\vec{k}) f_{\rho \sigma}^{\delta \vec{n}_1 , \delta \vec{n}_2}(P_0) \\
 & = \sum_{\vec{k}} \tilde{V}^{\rho}(\vec{k}) \tilde{V}^{\sigma}(-\vec{k}) \ch_{\rho \sigma}(P_0) ,
\end{align}
where $\ch_{\rho \sigma}$ is the Hamiltonian part of the Hessian matrix:
\begin{align}
    \ch_{\rho \sigma}(P_0) \equiv  \sum_{\delta \vec{n}_1} \sum_{\delta \vec{n}_2} e^{i \mu \vec{k} \cdot (\delta \vec{n}_1 - \delta \vec{n}_2)} f_{\rho \sigma}^{\delta \vec{n}_1 , \delta \vec{n}_2}(P_0).
\end{align}

Assuming the total number of time steps $N+1$ to be odd, we use the discrete Fourier transform for the time direction as changing integration variables
\be         
\tilde{V}^{\rho}(\tau_j,\vec k) = \frac{1}{\sqrt{N+1}} \sum_{k_0=-N/2}^{N/2}\widetilde{V}^{\rho}(k_0,\vec{k})e^{-\frac{2\pi i}{N+1}k_0j},
\ee
where $j=1,\cdots,N+1$ labels the time steps. When we take the time continuum limit $N \to \infty$ in the time interval $[0,T]$, we rescale $\widetilde{V}^\rho$ and use the following Fourier transformation
\be
\tilde{V}^{\rho}(\tau,\vec k)=\frac{1}{\sqrt{T}}\sum_{k_0=-\infty}^\infty\,\widetilde{V}^{\rho}(k_0,\vec{k})e^{-\frac{2\pi i}{T}k_0\tau}
\ee
for changing integration variables. The reality of the perturbations $V^\rho(\t,v)$ implies 
\be 
\widetilde{V}^{\rho}(-k_0,-\vec{k})=\widetilde{V}^{\rho}(k_0,\vec{k})^*.
\ee
Changing variables by Fourier transformation has the convenience of making the Hessian matrix block diagonal.

\subsection{Hessian matrix}\label{Hessian matrix}

When we expand the action $S$ in terms of perturbations $V^\rho$, the 0-th order vanishes due to the flat background, and the 1st order vanishes due to the equations of motion. We denote the quadratic order by $S^{(2)}$. When we express the quadratic action in terms of $\widetilde{V}^\rho(k_0,\vec{k})$, we take $N\to\infty$ and approximate $\sum_i\Delta\tau(\cdots ) \simeq \int_0^T\rmd \t(\cdots)$, and we use the normalization
\be
\frac{1}{T}\int_0^T\rmd \t \, e^{-\frac{2\pi i}{T}(k_0-k_0')\tau}=\delta_{k_0,k'_0}.
\ee

In our perturbative computation of one-loop effective action, we only turn on the perturbations $V^\rho$ of holonomies and fluxes, but we do not consider the perturbation of $h_v$, although $h_v$ are integration variables in the path integral formula. The reason is that the integration of $h_v$ only contribute to the definition of the initial state $\Psi_{[g']}$, and $h_v$ is only defined on the initial slice. Recall the discussion in Section \ref{Effective action and coherent states}. The holonomies and fluxes in our context correspond to the dynamical field $\phi$ in Section \ref{Effective action and coherent states}, while $h_v$ and their integrals correspond to parts of the initial state $\theta_+$. For instance in \eqref{eq:Wtheta}, the external source $J$ only couples to the dynamical field, whereas $J$ does not couple to the initial state $\theta^+$. In the following computation of the Hessian and one-loop determinant, we only perform the integration of the dynamical field $V^\rho$, while leaving the integration of $h_v$ untouched. 

Additionally, we do not perform the integration of the zero mode $V^\rho(k_0=0,\vec{k})$, because the zero mode results in the IR divergence of the one-loop effective action, as to be seen in a moment. Leaving the zero mode not integrated is an IR regularization in the quantum effective action.

Expanding the action $S$ in $V^\rho$ and truncating to quadratic order, we obtain the quadratic action
\begin{align}\label{quadraticaction000}
    S^{(2)} =& \sum_{k_0} \sum_{\vec{k}} \widetilde{V}_{\rho}(-k_0,-\vec k) [M_{VV}]^{\rho}{}_{\sigma}(k_0,\vec{k})\widetilde{V}^{\sigma}(k_0,\vec k).
\end{align}

The one-loop effective action relates to the determinant of the Hessian matrix by \eqref{eq:EA}. By \eqref{quadraticaction000}, the Hessian matrix is block-diagonal with each $18\times 18$ block $M_{VV}$ associated with fixed $k_0,\vec{k}$. The detailed form of $M_{VV}$ and relevant code can be downloaded in \cite{github}.

It turns out $DM \equiv \det M_{VV}$ satisfies the following transformations
\begin{align}\label{eq:symmetry_DM}
    DM(k_0, m_1,m_2,m_3) =& DM( k_0, m_2,m_3,m_1)= DM( k_0, m_3,m_1,m_2)\\
    DM(k_0, m_1,m_2,m_3) =& DM( - k_0, m_2,m_1,m_3) = DM( - k_0, - m_1,m_2,m_3) \\
    =& DM( - k_0, m_1, -m_2,m_3)= DM( - k_0, m_1,m_2,- m_3),
\end{align}
where $DM(k_0, m_1,m_2,m_3) \equiv DM\left(k_0, \vec{k} = \frac{2 \pi}{L} (m_1,m_2,m_3)\right)$.
The first transformation is nothing else but the lattice rotation symmetry, while the second implies that $DM$ is invariant under the time-reversal and parity transformation.
Moreover, $DM(k_0, m_1,m_2,m_3)$ is given by the following expression 
\begin{align}
    DM(k_0, m_1,m_2,m_2) =& \frac{\left(\frac{2 \pi}{T}k_0 \right)^{18}\mu^{36}} {a^{36} \beta^{18}} \prod_{i=1}^{6} \left(1 - \frac{\alpha_{i,\vec{m}}}{\left(\frac{2 \pi}{T}k_0 \right) \sqrt{P_0}} \right)\left(1 - \frac{\overline{\alpha_{i,\vec{m}}} }{\left(\frac{2 \pi}{T}k_0 \right) \sqrt{P_0}} \right). \label{eq:general_DM}
\end{align}
Here $\alpha_i$ as complex functions of $\vec{m}$ have complicated analytical expressions. There are special cases for $\vec{m}$ when $\vec{m}$ lies on the axes or planes of the momentum lattice, in which case $DM$ becomes 
\begin{align}\label{eq:special_DM}
    DM(k_0, m_1,m_2,0) =& \frac{\left(\frac{2 \pi}{T}k_0 \right)^{18}\mu^{36}} {a^{36} \beta^{18}} \prod_{i=1}^{j} \left(1 - \frac{\tilde{\alpha}_{i,\vec{m}}}{\left(\frac{2 \pi}{T}k_0 \right)^2 P_0} \right)\left(1 - \frac{\overline{\tilde{\alpha}_{i,\vec{m}}}}{\left(\frac{2 \pi}{T}k_0 \right)^2 P_0} \right)  \prod_{i=1}^{6-2j} \left(1 - \frac{\beta_{i,\vec{m}}}{\left(\frac{2 \pi}{T}k_0 \right)^2 P_0} \right),
\end{align}
where $\b_i$ are real functions of $\vec{m}$, and 
\begin{align}\label{eq:special_DM1}
       DM(k_0, m_1,0,0) =& \frac{\left(\frac{2 \pi}{T}k_0 \right)^{18}\mu^{36}} {a^{36} \beta^{18}} \left(1 - \frac{\alpha_{m_1}}{\left(\frac{2 \pi}{T}k_0 \right)^2 P_0} \right)^2 , \\
       \alpha =&-(\sin ^2(|k| \mu ) \left(1 - \left(\beta ^2+1\right) \cos (|k| \mu )-\beta ^2\right) ) \label{alphakmu}
\end{align}
We notice that the expression of $DM$ closely relates to the propagators. For $DM(k_0, m_1,0,0) $, the propagator corresponds to $2$ non-trivial degrees of freedom, while for $DM(k_0, m_1,m_2,0) $ and $DM(k_0, m_1,m_2,m_3) $ we have $6$ non-trivial degrees of freedom. These $6$ degrees of freedom are the full number degrees of freedom we have in the deparametrized model, encoded in $9$ canonical pairs modulus $3$ Gauss constraints. Moreover, for $DM(k_0, m_1,m_2,m_3) $ we do not recover the form of the standard scalar propagator, and the time-reversal symmetry $DM(k_0) = DM(-k_0)$ is broken, similar to the spinor field. However, from the symmetry of $DM$, we see that the propagator is invariant under the time-reversal and parity transformation.

When we consider the regime $\mu k\ll1$ and ignore higher orders in $\mu k$, or equivalently, $\mu\to0$ with $k_0,\vec{k}$ fixed, the expression of $DM$ is simplified drastically:
\begin{align}\label{lowEDM}
  DM(k_0,\vec{k}) \simeq \frac{\left(\frac{2 \pi}{T}k_0 \right)^{18}\mu^{36}} {a^{36} \beta^{18}} \left(1 - \frac{|\vec{k}|^2}{\left(\frac{2 \pi}{T}k_0 \right)^2 P_0} \right)^2, %
\end{align}
where we recover the time-reversal symmetry and full SO(3) rotational invariance. The Hessian matrix is dimensionless, as $k_0 \sim |\vec{k}| \sim [{\rm length}]^{-1}$ , $a \sim \mu \sim \lambda \sim [{\rm length}]^1$.

The above results indicates that $DM=0$ at $k_0=0$. The one-loop effective action relates to the logarithm of $DM$ and thus is divergent at $k=0$. This is the reason why we make the IR regularization by not taking into account the integration of zero mode $V^\rho(k_0=0,\vec{k})$.

\section{Propagator}\label{sec:5}
Clearly the poles of the propagators can be read from the inversus of \eqref{eq:general_DM}, namely
\begin{align}
    P(k_0,\vec{m}) = \frac{1}{DM(k_0,\vec{m})} = \frac{a^{36} \beta^{18}}{\mu^{36}}  \frac{1}{ \left(\frac{2 \pi}{T}k_0 \right)^{6} \prod_{i=1}^{6} \left(\left(\frac{2 \pi}{T}k_0 \right) - \frac{\alpha_i}{ \sqrt{P_0}} \right)\left(\left(\frac{2 \pi}{T}k_0 \right)  - \frac{\overline{\alpha_i} }{ \sqrt{P_0}} \right)} , \quad \alpha_i \in \mathbb{C}
\end{align}
As we mentioned before, these corresponds to $3$ unphysical degrees of freedom from Gauss constraints and $6$ physical degrees of freedom. In this section, we will investigate more in detail the form of the propagator. In general, the poles of the physical degrees of freedom relates to the complex number $\alpha_i$, except in the special case where $\vec{k}$ lies along the axes, e.g. $\vec{k}=k_x$. In such a case we have two real poles which can be given analytically as in shown in \eqref{eq:special_DM}.

\begin{figure}
    \centering
    \includegraphics[width=0.46\linewidth]{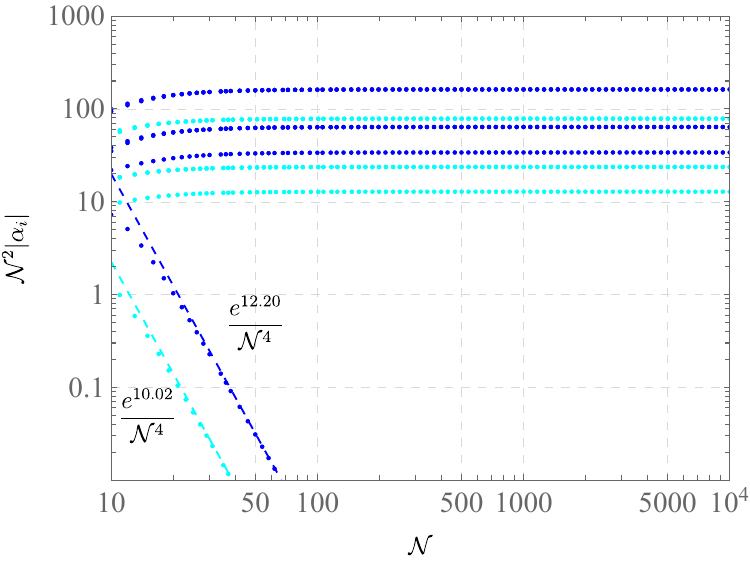}
    \includegraphics[width=0.46\linewidth]{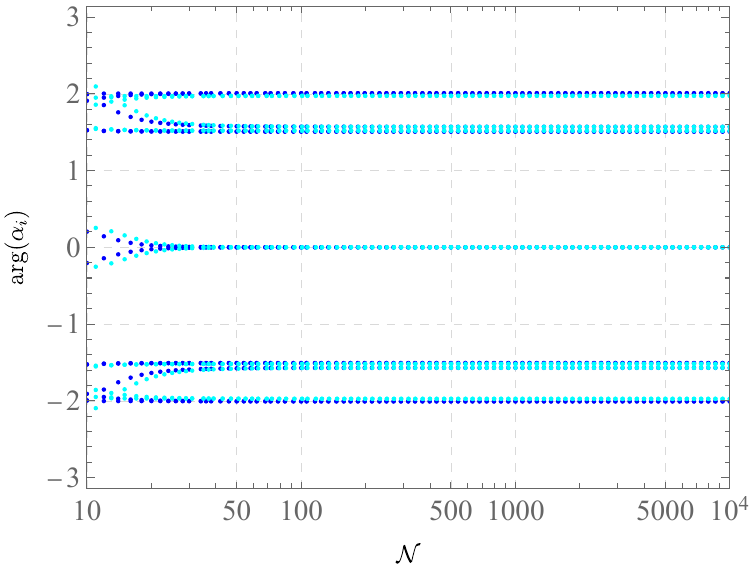}

    \begin{tabular}{|c|c|c|c|c|c|c|}
        \hline
    \multicolumn{7}{|c|}{$\mathcal{N}^2|\alpha_i|$}   \\\hline
         $\mathcal{N} = 10^8$ & 23.7 & 23.7  &\num{2.249e-28}  & 78.557 & 78.557 & 12.823\\
        $\mathcal{N} = 10^8+1$ & 63.861 & 63.861 & \num{1.994e-27} & 162.211 & 162.211 & 33.931 \\ \hline 
     \multicolumn{7}{|c|}{$\arg(\alpha_i)/\pi$}  \\  \hline
         $\mathcal{N} = 10^8$ & 0.639 & 0.639 & 0.5 & 0.480 & 0.480 & \num{1.351e-29}\\
        $\mathcal{N} = 10^8+1$ & 0.649 & 0.649 & 0.5  & 0.483 & 0.483 & \num{4.135e-30} \\
        \hline
    \end{tabular}

    \caption{Left: Absolute values of $\mathcal{N}^2 \alpha_i$ as a function of $\mathcal{N}$. Right: Arguments of $\alpha_i$ of a given $\vec{k}$ versus the total number of vertices along each direction $\mathcal{N}$. The $\vec{k}$ is given in the following way. For $\mathcal{N}$ odd (cyan points), $\vec{k} = \frac{2\pi}{L}\left( \frac{\mathcal{N} -1}{2}-2, \frac{\mathcal{N} -1}{2}-1, \frac{\mathcal{N} -1}{2} \right)$, for $\mathcal{N}$ even (blue points), $\vec{k} = \frac{2\pi}{L}\left( \frac{\mathcal{N}}{2} -3, \frac{\mathcal{N}}{2} -2, \frac{\mathcal{N}}{2} -1 \right)$. For both cases there will be 5 $\mathcal{N}^2 \alpha_i$ converging to 3 different complex values and 1 $\mathcal{N}^2  \alpha_i$ converging to $0$ asymptotically with power $\mathcal{N}^{-4} $. However, the converging value is different for odd and even cases. The values at $\mathcal{N}=10^8$ and $\mathcal{N}=10^8+1$ are given in the table.}
    \label{fig:kmax-1-2-3}
\end{figure}

\begin{figure}
    \centering
    \includegraphics[width=0.46\linewidth]{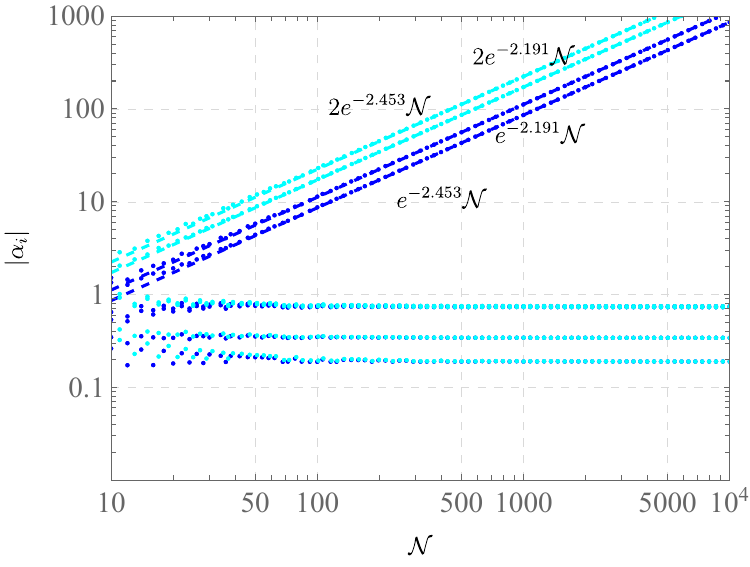}
    \includegraphics[width=0.46\linewidth]{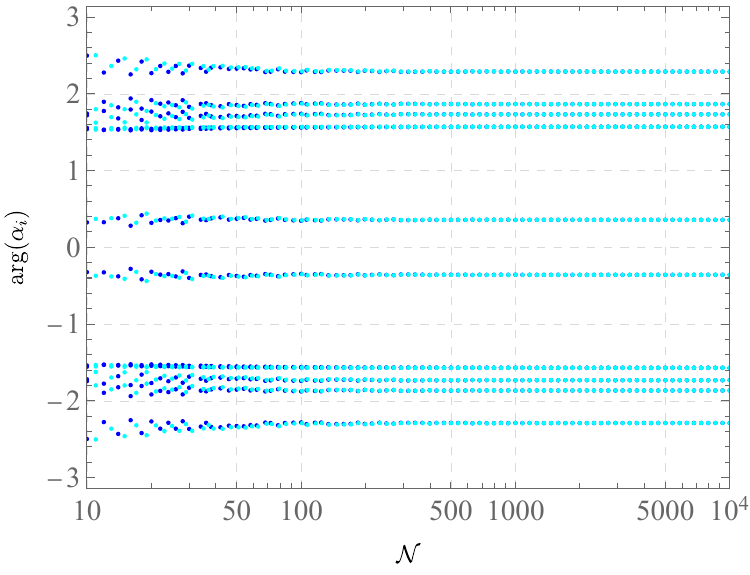}

    \begin{tabular}{|c|c|c|c|c|c|c|}
        \hline
    \multicolumn{7}{|c|}{$|\alpha_i|$}   \\\hline
         $\mathcal{N} = 10^8$ & 0.73 & 0.19 & 0.748 & \num{1.118e7}  & \num{8.6e7} & 0.342\\
        $\mathcal{N} = 10^8+1$ & 0.365 &0.095 &0.374 &\num{1.118e7} &\num{8.6e7} &0.171 \\ \hline 
     \multicolumn{7}{|c|}{$\arg(\alpha_i)$}  \\  \hline
         $\mathcal{N} = 10^8$ & 0.594 & 0.7288 & 0.5516 & 0.5 &0.5 & 0.114\\
        $\mathcal{N} = 10^8+1$ & 0.594 & 0.7288 & 0.5516 & 0.5 &0.5 & 0.114 \\
        \hline
    \end{tabular}

    \caption{Left: Absolute values of $ \alpha_i$ as a function of $\mathcal{N}$. Right: Arguments of $\alpha_i$ of a given $\vec{k}$ versus the total number of vertices along each direction $\mathcal{N}$. The $\vec{k}$ is given in the following way. For $\mathcal{N}$ odd, $\vec{k} = \frac{2\pi}{L}\frac{\mathcal{N} -1}{2}\left( \frac{1}{3} , \frac{1}{2} , 1\right)$, for $\mathcal{N}$ even, $\vec{k} = \frac{2\pi}{L}\left( \frac{\mathcal{N}}{6}, \frac{\mathcal{N}}{4}, \frac{\mathcal{N}}{2} -1 \right)$. In the plot, the $|\alpha_i|$ corresponding to the odd $\mathcal{N}$ (blue points) is scaled with $2$ compare to the $|\alpha_i|$ for the even $\mathcal{N}$ (cyan points). For both cases there will be 4 $\alpha_i$ converging to 4 different complex values where the converging value for the odd configuration is exactly twice the converging value of the even configuration. Furthermore there will be  2 $\alpha_i$ which scale as $|\vec{k}| \sim \mathcal{N}$, which have the same scaling as the $2$ non-trivial tensor modes in the classical limit.}
    \label{fig:kmax1d2d3}
\end{figure}

\begin{figure}
    \centering
    \includegraphics[width=0.46\linewidth]{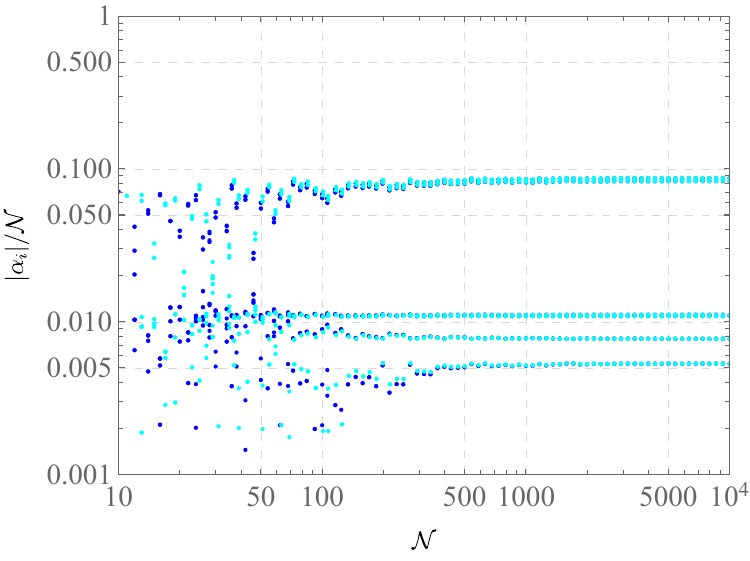}
    \includegraphics[width=0.46\linewidth]{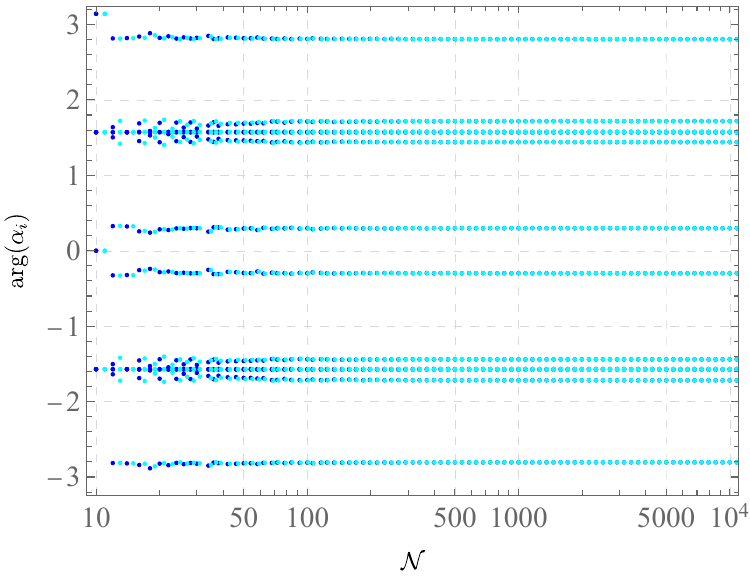}

    \begin{tabular}{|c|c|c|c|c|c|c|}
        \hline
    \multicolumn{7}{|c|}{$|\alpha_i|/\mathcal{N}$}   \\\hline
         $\mathcal{N} = 10^8$ & 0.03 & 0.032 & 0.185 & 0.19 & 0.029 & 0.031\\
        $\mathcal{N} = 10^8+1$ & 0.03 & 0.032 & 0.185 & 0.19 & 0.029 & 0.031 \\ \hline 
     \multicolumn{7}{|c|}{$\arg(\alpha_i)$}  \\  \hline
         $\mathcal{N} = 10^8$ & 0.893 & 0.547 & 0.502 & 0.499 & 0.458 & 0.095\\
        $\mathcal{N} = 10^8+1$ & 0.893 & 0.547 & 0.502 & 0.499 & 0.458 & 0.095 \\
        \hline
    \end{tabular}

    \caption{Left: Absolute values of $\frac{\alpha_i}{\mathcal{N}}$ as a function of $\mathcal{N}$. Right: Arguments of $\alpha_i$ of a given $\vec{k}$ versus the total number of vertices along each direction $\mathcal{N}$. The $\vec{k}$ is given in the following way. For $\mathcal{N}$ odd (cyan points), $\vec{k} = \frac{2\pi}{L}\frac{\mathcal{N} -1}{2}\left( \frac{1}{6}, \frac{1}{4}, \frac{1}{2} \right)$, for $\mathcal{N}$ even (blue points), $\vec{k} = \frac{2\pi}{L}\frac{\mathcal{N}}{2}\left( \frac{1}{6}, \frac{1}{4}, \frac{1}{2} \right)$. The $\frac{\alpha_i}{\mathcal{N}}$ will converging to 5 different complex values for both cases. The even and odd cases converging to the same value and all the $6$ modes have the same scaling (scaled as $|\vec{k}| \sim \mathcal{N}$) as non-trivial modes in the classical limit. }
    \label{fig:kmax2d4d6}
\end{figure}

\begin{figure}
    \centering
    \includegraphics[width=0.46\linewidth]{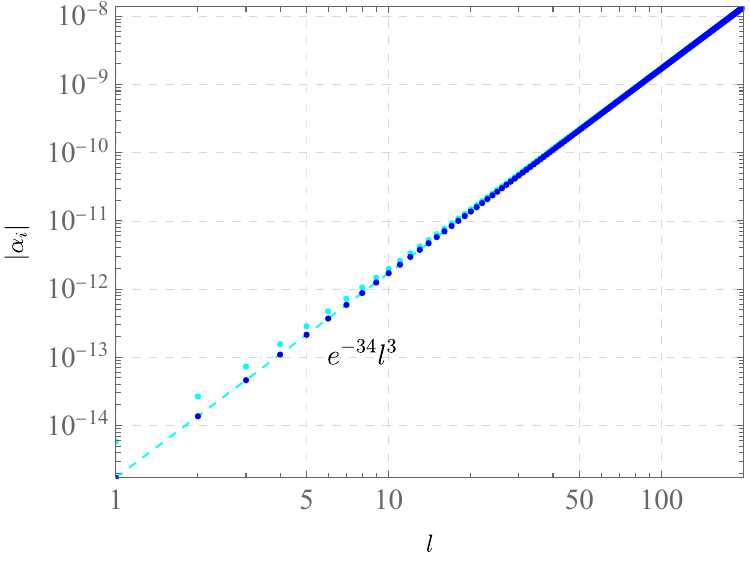}
    \includegraphics[width=0.46\linewidth]{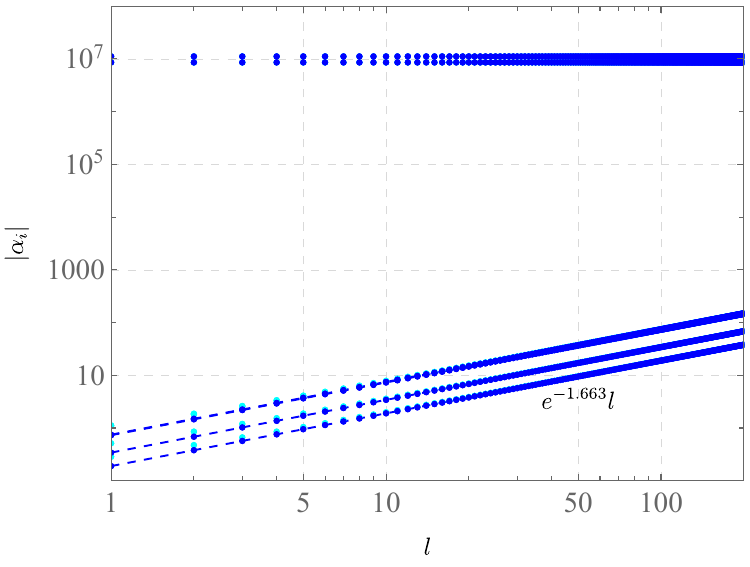}

    \caption{Plots of the absolute values of $\alpha_i$ for $\vec{k}$'s with a given $\mathcal{N} = 10^8$ as a function of the difference parameter $l$, which we subtract from $\mathcal{N}$. The $\vec{k}$ is given in the following way. Left: For $\mathcal{N}$ odd (cyan points), $\vec{k} = \frac{2\pi}{L}\left( \frac{\mathcal{N} -1}{2} -l \right) \left(1,1,1 \right)$, for $\mathcal{N}$ even (blue points), $\vec{k} = \frac{2\pi}{L}\left( \frac{\mathcal{N}}{2} -l \right) \left(1,1,1 \right)$. Right: For $\mathcal{N}$ odd (cyan points), $\vec{k} = \frac{2\pi}{L}\left(\frac{\mathcal{N} -1}{6} -l , \frac{\mathcal{N} -1}{4} -l,\frac{\mathcal{N} -1}{2} -l  \right)$, for $\mathcal{N}$ even (blue points), $\vec{k} = \frac{2\pi}{L}\left(\frac{\mathcal{N}}{6} -l , \frac{\mathcal{N} }{4} -l,\frac{\mathcal{N}}{2} -l  \right)$. In the first plot (difference from $(\mathcal{N}-1)/2$ in all directions) there is a scaling like $l^3$, while in the second (difference from different  values in each direction)  it scales linearly $l$. }
    \label{fig:kmax-scale}
\end{figure}

Notice that for a fixed region of space with fixed length $L$ along each direction, an arbitrarily large but finite $\mathcal{N}$ will always give a high energy cutoff to the momentum $\vec{k}$, as $|k| \leq k_{\text{max},\mathcal{N}} \equiv \frac{\pi}{\mu} = \frac{\pi \mathcal{N}}{L}$. As a result, for the $\vec{k}$ configurations that have components close to $k_{\text{max},\mathcal{N}}$, we expect it not to converge to its classical value under the limit $\mathcal{N} \to \infty$, which is not regularized. In contrast, for $|k| \leq k_{\text{max},\mathcal{N}}$, especially for a given fixed $\vec{k}$ that does not scale with $\mathcal{N}$, we expect the classical values under $\mathcal{N} \to \infty$. Here we give some numerical results of the poles $\alpha_i$ for different configurations of $\vec{k}$, especially in the lattice continuum limit. In Figures (\ref{fig:kmax-1-2-3}-\ref{fig:kmax-scale}) we show how different configurations of $\vec{k}$, chosen to be scaled by $\mathcal{N}$, converge (or scale) under the limit $\mathcal{N} \to \infty$. In all the numerical examples we set $L=2 \pi$ and $a=\beta
=1$. Here we set $P_0=1$ as well. It turns out that one can actually estimate a convergence (or scaling) of any configurations even when the $\vec{k}$ are scaled by $\mathcal{N}$.

In order to extract more easily the properties of the propagator, we will integrate out the momentum perturbations $\mathcal{X}$. The integration leads to the following Hessian matrix for  $\mathcal{Y}$ only
\begin{align}
    \widetilde{M}_{\mathcal{Y}\mathcal{Y}} = {M}_{\mathcal{Y}\mathcal{Y}} - {M}_{\mathcal{Y}\mathcal{X}} ({M}_{\mathcal{X}\mathcal{X}})^{-1} {M}_{\mathcal{X}\mathcal{Y}}
\end{align}
where ${M}_{\mathcal{Y}\mathcal{Y}}$ denotes the diagonal block corresponding $V^{1} \cdots V^{9}$ in $M_{VV}$ while $\mathcal{X}$ refers to the block with $V^{10} \cdots V^{18}$. The determiant of $M_{VV}$ now is given by 
\begin{align}
    \det M_{VV} = \det \widetilde{M}_{\mathcal{Y}\mathcal{Y}} \det {M}_{\mathcal{X}\mathcal{X}}
\end{align}
where we have in general now  that
\begin{align}
    \det {M}_{\mathcal{X}\mathcal{X}}[k_0, \vec{m}] = \alpha_{\vec{m}} \left( \sqrt{P_0} \mu \right)^9, \qquad \alpha_{\vec{m}} \in \mathbb{R} \;.
\end{align}
This is independent of $k_0$, but depends on $\vec{m}$. Now, elements in $\widetilde{M}_{\mathcal{Y}\mathcal{Y}} $ contain terms quadratic in $k_0$. Its eigenvalue is given by the solutions of the characteristic equation
\begin{align}\label{eq:eigen_propa}
    \det(\widetilde{M}_{\mathcal{Y}\mathcal{Y}}  - \lambda \mathbb{I}_9 ) = 0 .
\end{align}
The eigenvalue equation turns out to be a complicated equations which can only be solved numerically. However, we find that, although $P(k_0,\vec{m})$ appears to have a good separation between different modes, the eigenvalues do not. For a general $\vec{m}$, \eqref{eq:eigen_propa} has $9$ distinct solutions. For $\vec{m}$ in the from of $\vec{m} = (m_1,0,0)$ (and its permutation or parity transformation), we have $6$ distinct eigenvalues, which appears as $3$ eigenvalues with degeneracy $2$ and $3$ non degenerate eigenvalues. Moreover, all the eigenvalues scales as $k_0^2$ in the limit $k_0 \to \infty$, which gives in total $k_0^{-18}$ for the propagator. However, in the limit $k_0 \to 0$, different patterns of $\vec{m}$ has different scalings, which is summarized in Table \ref{table_scaling_propa}. It is easy to check that the overall scaling is compatible with the minimal powers of $k_0$ which appears in $\det M_{VV}$ as given in \eqref{eq:general_DM} and \eqref{eq:special_DM}.
\begin{table}[h!]
    \centering
    \begin{tabular}{|c|c|c|}
        \hline
        $\vec{m}$ & scaling & overall scaling\\
        \hline
        $(m_1,0,0)$ & 5 eigs. with $k_0^2$, 1 eigs. with $k_0^4$, 3 eigs with $1$ & $k_0^{14}$\\
        $(m_1,m_2,0)$ & 4 eigs. with $k_0$, 1 eigs. with $k_0^2$, 4 eigs with $1$ & $k_0^6$\\
        $(m_1,m_2,m_3)$ & 3 eigs. with $k_0^2$, 6 eigs with $1$ & $k_0^6$\\
        \hline
    \end{tabular}
    \caption{\label{table_scaling_propa}Scaling of $k_0$ for the eigenvalues of $\widetilde{M}_{\mathcal{Y}\mathcal{Y}}$ in the limit $k_0 \to 0$.}
\end{table}
As a summary, in general there does not exist a change of variables such that one can separate different modes appearing in the full propagator. Moreover, there does not exist a change of variables in which one could separate the gauge degrees of freedom corresponding to the Gauss constraint from the physical degrees of freedom, as this would correspond to a block diagonalization with 3 eigenvalues given by $\sim k_0^2$. As a result, if there exists such change of variables, this will correspond to a different theory, as it will lead to a different effective action.

\begin{figure}
    \centering
    \includegraphics[width=0.31\linewidth]{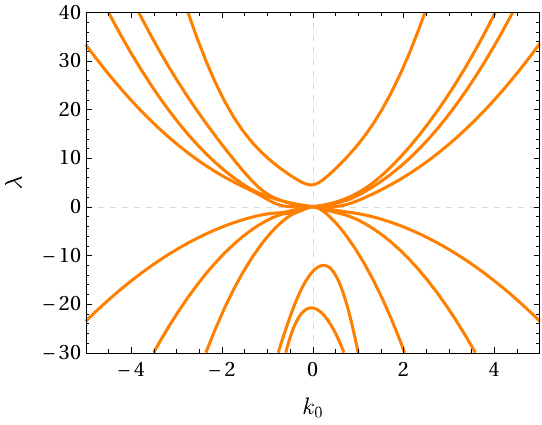}
    \includegraphics[width=0.31\linewidth]{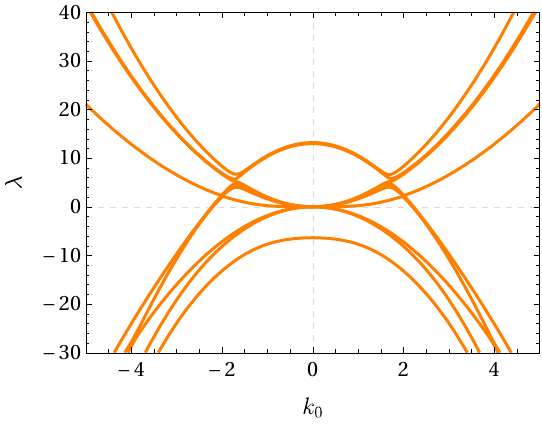}
    \includegraphics[width=0.31\linewidth]{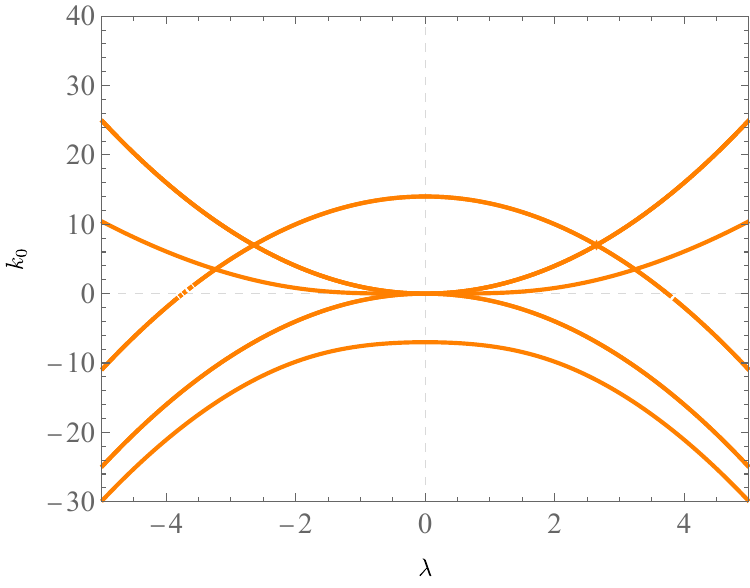}
    \caption{The plots 
 of eigenvalues $\lambda(k_0)$ of $\widetilde{M}_{\mathcal{Y}\mathcal{Y}}$ with fixed momentum $\vec{k}=\frac{2\pi}{L}{(1,2,3)}$ as a function of time components $k_0$ for different $\mathcal{N}$: left: $\mathcal{N} = 11$, center: $\mathcal{N} = 23$, right: $\mathcal{N} = 10^6$. From the right plot we obtain a same results as the classical limit, where the real roots of $\lambda$ happens at $0$ and $\pm \sqrt{1^2+2^2+3^2} \simeq 3.741$. }
    \label{fig:eig_k123}
\end{figure}

\begin{figure}
    \centering
    \includegraphics[width=0.31\linewidth]{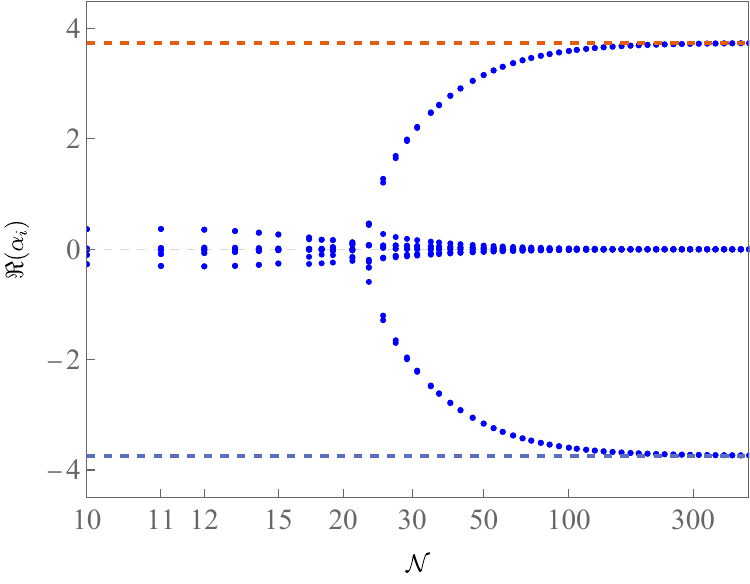}
    \includegraphics[width=0.31\linewidth]{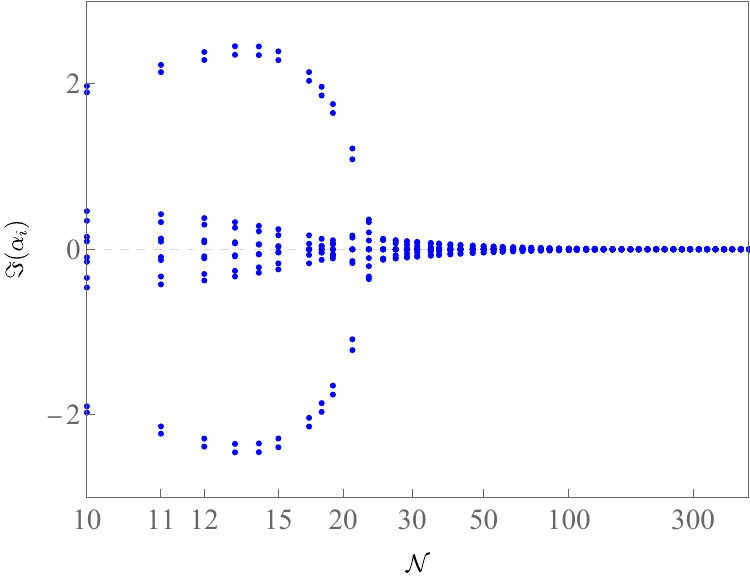}
    \includegraphics[width=0.31\linewidth]{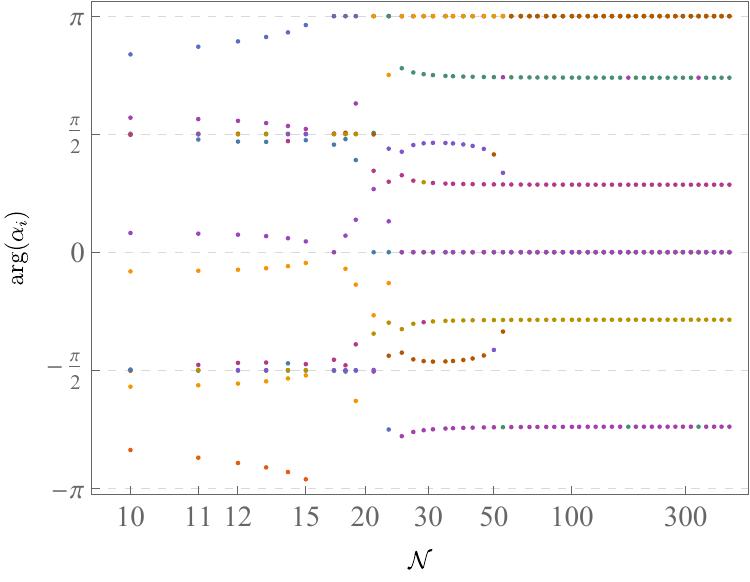}
    \caption{The plots 
 of $\alpha_i$ with fixed momentum $\vec{k}=\frac{2\pi}{L}{(1,2,3)}$ versus $\mathcal{N}$. It is clear that $\alpha_i$ converge to the classical limit, where we have two $\alpha_i$ converging to $\pm \sqrt{1^2+2^2+3^2} \simeq 3.741$ corresponding to the tensor modes, two converging to $0$ corresponding to the vector modes and two converging to $0$ along the complex plane, corresponding to the scalar modes. These can be found in  \eqref{eq:vector_and_tensor} and \eqref{eq:scalar}, respectively. }
    \label{fig:alpha_k123}
\end{figure}

Finally, we remark that in the continuum limit the inverse of the eigenvalues of $\widetilde{M}_{\mathcal{Y}\mathcal{Y}}$ are given by the following sets
\begin{align}
   &\left\{ - \frac{a^2 \sqrt{P_0}}{\left(\frac{2 \pi}{T}k_0 \right)^2} , - \frac{a^2 \sqrt{P_0}}{\left(\frac{2 \pi}{T}k_0 \right)^2}, - \frac{a^2 \sqrt{P_0}}{\left(\frac{2 \pi}{T}k_0 \right)^2} \right\}, \label{eq:gauss_mode}\\
   &\left\{ \frac{a^2 \sqrt{P_0}}{\left(\frac{2 \pi}{T}k_0 \right)^2},  \frac{a^2 \sqrt{P_0}}{\left(\frac{2 \pi}{T}k_0 \right)^2} \right\}, \qquad 
   \left\{ \frac{a^2 \sqrt{P_0}}{\left(\frac{2 \pi}{T}k_0 \right)^2 - \frac{\vec{k}^2}{P_0} },  \frac{a^2 \sqrt{P_0}}{\left(\frac{2 \pi}{T}k_0 \right)^2 - \frac{\vec{k}^2}{P_0} } \right\} \label{eq:vector_and_tensor}\\
     &\left\{  \frac{a^2 \sqrt{P_0}}{\left(\frac{2 \pi}{T}k_0 \right)^2 + \frac{\vec{k}^2}{P_0} + \sqrt{8 \left(\frac{2 \pi}{T}k_0 \right)^4 + \left( \left(\frac{2 \pi}{T}k_0 \right)^2 + \frac{\vec{k}^2}{P_0} \right)^2 } },   \frac{a^2 \sqrt{P_0}}{\left(\frac{2 \pi}{T}k_0 \right)^2 + \frac{\vec{k}^2}{P_0} - \sqrt{8 \left(\frac{2 \pi}{T}k_0 \right)^4 + \left( \left(\frac{2 \pi}{T}k_0 \right)^2 + \frac{\vec{k}^2}{P_0} \right)^2 } }\right\} \label{eq:scalar}
\end{align}
These are the propagators in the continuum limit. Denoting $\l^\alpha(k_0,\vec{k})$ ($\a=1,\cdots,9$) the eigenvalues of $\widetilde{M}_{\cy\cy}$ (the inverse of \eqref{eq:gauss_mode} - \eqref{eq:scalar}), the eigenvalues multiplying the corresponding eigenvectors $\widetilde{\psi}^\alpha(k_0,\vec{k})$ are: 
\be
\l^\alpha(k_0,\vec{k})\, \widetilde{\psi}^\alpha(k_0,\vec{k})=0,\qquad \forall \a=1,\cdots,9,\quad \text{ (no sum in $\a$)}
\ee
They are equivalent to the linearized equations of motion. The eigenvectors $\widetilde{\psi}^\alpha$ are certain linear combinations of the perturbations $\widetilde{V}^\rho$. The convergence of the general $\widetilde{M}_{\mathcal{Y}\mathcal{Y}}$ to its continuum limit when $\mu \to 0$ is shown in Figure \ref{fig:eig_k123} and \ref{fig:alpha_k123}.

The first $3$ eigenvalues of $\widetilde{M}_{\cy\cy}$ from \eqref{eq:gauss_mode} correspond to the eigenvectors given by {$\mathcal{Y}^{a}(e_{I}) - \mathcal{Y}^{I}(e_{a}), a \neq I$}. The linearized equations of motion imply these eigenvectors vanish except at $k_0=0$, so their inverse Fourier transformations are conserved in the $\tau$-evolution. Vanishing eigenvectors $\mathcal{Y}^{a}(e_{I}) - \mathcal{Y}^{I}(e_{a})=0$ is equivalent to the linearized Gauss constraint. Indeed,  linearizing the continuous Gauss constraint $\epsilon_{ijk}E^a_iK^j_a=0$ at the background with $K_0=0$ gives $K^j_a$ is symmetric which implies the perturbation of $E^a_i$ is symmetric, since the extrinsic curvature $K_{ab}$ is symmetric. Therefore, the linearized equations of motion for \eqref{eq:gauss_mode} are linearizing the conservation of the Gauss constraint function $G$, while the constraint $G=0$ is imposed on the initial slice by the $h$ integral of the path integral formula.

The first two inverse eigenvalues in \eqref{eq:vector_and_tensor} correspond to vector modes, whose eigenvectors are given by, e.g. $\mathcal{Y}^{1}(e_{2}) + \mathcal{Y}^{2}(e_{1})$ and  $\mathcal{Y}^{1}(e_{3}) + \mathcal{Y}^{3}(e_{1})$ in the case $\vec{k} = (k_x,0,0)$. The last two inverse eigenvalues in \eqref{eq:vector_and_tensor} are tensor modes, whose eigenvectors are given by, e.g. $\mathcal{Y}^{2}(e_{3}) + \mathcal{Y}^{3}(e_{2})$ and  $\mathcal{Y}^{2}(e_{2}) -\mathcal{Y}^{3}(e_{3})$ in the case $\vec{k} = (k_x,0,0)$. Finally, the inverse eigenvalues in \eqref{eq:scalar} correspond to the scalar modes whose eigenvectors are  $\mathcal{Y}^{1}(e_{1})$ and $\mathcal{Y}^{2}(e_{2}) + \mathcal{Y}^{3}(e_{3}) $ in the case of $\vec{k} = (k_x,0,0)$. {The convergence of the eigenvectors of the general $\widetilde{M}_{\mathcal{Y}\mathcal{Y}}$ can be verified numerically, where for fixed $\vec{k}=\frac{2\pi}{L}(1,2,3)$ and $L=2 \pi$ we find the difference with the continuum limit is at the order of $\mathcal{N}^{-1}$.  }

The poles of \eqref{eq:vector_and_tensor} - \eqref{eq:scalar} are given by $k_0=0$ except for the last two in \eqref{eq:vector_and_tensor} for tensor modes. Correspondingly, the inverse Fourier transformations of their eigenvectors are conserved in time $\t$, except for the tensor modes. This is consistent with the fact that the tensor modes are the only dynamical degrees of freedom corresponding to the propagating graviton, while other eigenvectors are degrees of freedom of the dust co-moving with the reference frame for $(\tau,\vec{\sigma})$.

\section{One-loop effective action in long wavelength approximation}\label{1-loop effective action in long wavelength approximation}

\subsection{The Gaussian functional determinant}

Recalling eq. \eqref{eq:EA}, one has to evaluate the Hessian $H$ and compute its determinant. In a QFT setting, under the path integral, computing the functional determinant  amounts to perform the Gaussian integral over the fluctuation. In a discretized setting instead, this procedure amounts to diagonalize the operator $H$ and to multiply the eigenvalues. In particular, let us consider as an example a particle in a well of depth $A$ and lenght $L$. The eigenvalues are $\lambda_n = \frac{n^2\pi^2}{L^2}+A, \quad n \in \mathbb{N}\setminus \{0\}$. We need to compute
\begin{equation}\label{eq:detwell}
    \frac{\text{det} \left(-\frac{d^2}{dx^2}+A\right)}{\text{det} \left(-\frac{d^2}{dx^2}\right)} = \prod_{n = 1}^{+\infty} \frac{\frac{n^2\pi^2}{L^2}+A}{\frac{n^2\pi^2}{L^2}} = \prod_{n = 1}^{+\infty} \left( 1+ \frac{L^2A}{n^2 \pi^2}\right) = \frac{\sinh(L\sqrt{A})}{L\sqrt{A}}
\end{equation}
Note that we took the quotient as not to have to bother with the uninteresting divergent constant.
It is also important to emphasize that here we considered a Euclidean toy model. In order to compute the Lorentzian counterpart, a Wick rotation can be performed, leading to $\frac{\sin(L\sqrt{A})}{L\sqrt{A}}$.

In our context, the determinant of Hessian is given by product of $DM$ computed in Section \ref{Hessian matrix} over all $\vec{k}$ and $k_0$. In this section, we insert the result of $DM$ in \eqref{lowEDM}. This result in \eqref{lowEDM} is valid only when $\mu |\vec{k}|\ll1$ and neglecting higher orders in $\mu |\vec k|$ in e.g. \eqref{alphakmu}. In other words, this result comes from the long wavelength approximation in the regime $|\vec k|^{-1}\gg \mu$. Computing the determinant beyond the long wavelength approximation is present in Section \ref{Beyond the long wavelength approximation}. Interestingly, the computations with and without the approximation give similar result for the effective action, while the computation with the approximation is much more explicit.  

The determinant of Hessian shares similarity with \eqref{eq:detwell}
\be  
\det\left(H\right)=\prod_{k_0\neq 0}\prod_{\vec{k}}DM(k_0,\vec{k})=\mathcal{C}\prod_{k_{0},\vec{k}}\left(1-\zeta^2\frac{|\vec{k}|^{2}}{k_{0}^{2}}\right)^{2},\qquad \zeta=\frac{T}{2\pi\sqrt{P_0}},
\ee
The only dependence of $\det\left(H\right)$ on the background field is the dependence on $P_0$. The field-independent overall constant $\mathcal{C}$ is 
\be 
\mathcal{C}&=&\prod_{k_{0},\vec{k}}\frac{\left(\frac{2 \pi}{T}k_0 \right)^{18}\mu^{36}} {a^{36} \beta^{18}}
\ee
whose logarithm only contributes a constant term to the effective action.

If we do not take into account the zero-mode corresponding to $k_0=0$, we can explicitly compute the product over $k_0$ as it is done for the quantum mechanical case in Eq.\eqref{eq:detwell}
\be 
\left[\prod_{k_0=1}^{\infty}\left(1-\zeta^{2}\frac{|\vec{k}|^{2}}{k_0^{2}}\right)^{2}\right]^{2}=\frac{\sin^{4}\left(\pi\zeta|\vec{k}|\right)}{\pi^{4}\zeta^{4}|\vec{k}|^{4}}
=\lt[\frac{\sin\left(\xi|\vec{m}|\right)}{\xi|\vec{m}|}\rt]^4,\qquad \xi=\frac{2\pi^{2}\zeta}{L}=\frac{\pi  T}{L\sqrt{P_{0}}}
\ee

The one-loop effective action $S_{1L}$ is a function of background field variables. In our case, the background field is a single variable $P_0$. This is similar to the local potential approximation in the discussion toward \eqref{Veffphi111}. Equivalently, $S_{1L}$ can be understood as function of $\xi$ \footnote{$\int d\widetilde{V}d\widetilde{V}^{*}\ e^{i\widetilde{V}^{\dagger}H\widetilde{V}}=\det(H)^{-1}=e^{-\log\det(H)}$, where the Hessian matrix $H$ is Hermitian.}
\be 
S_{1L}\left(\xi\right)=-\log\det (H)=-4\sum_{\vec{m}}\log\left[\frac{\sin\left(\xi|\vec{m}|\right)}{\xi|\vec{m}|}\right]+\text{const.}
\ee
Any constant term in the action can be neglected. The derivative of $S_{1L}$ is given by
\be 
\partial_{\xi}S_{1L}\left(\xi\right)=-4\sum_{\vec{m}}\left[|\vec{m}|\cot\left(\xi|\vec{m}|\right)-\frac{1}{\xi}\right] =-4\sum_{\vec{m}\neq 0}\lt[|\vec{m}|\cot\left(\xi|\vec{m}|\right)-\frac{1}{\xi}\rt].
\ee

$S_{1L}$ is divergent when $\xi|\vec{m}| \in \pi \mathbb{Z}\setminus\{0\}$ for some $\vec{m}$. But this divergence can be removed by introducing the standard
$i\epsilon$-regularization. The regularization is obtained by adding a phase to the time $T \to T e^{i\epsilon}$ with $\epsilon>0$. 

We assume the evolution time $T$ is very large such that $\xi\sim T/L\gg1$. By the $i\epsilon$-regularization,
\be
\cot\Big(\xi|\vec{m}|\left(1+i\epsilon\right)\Big)\simeq-i\left(1+2e^{-2 |\vec{m}| \xi  \epsilon +2 i |\vec{m}| \xi }\right),
\ee
for nonzero $|\vec{m}|$. A finite $\epsilon$ and $\xi\sim T/L\gg1$ make $e^{-2 |\vec{m}| \xi  \epsilon}$ suppressed exponentially. As a result,
\be 
\partial_{\xi}S_{1L}\left(\xi\right)\simeq 4if(\cn)+\frac{4}{\xi}\cn^3
\ee
where 
\be 
f(\cn)&=&\sum_{{m}_1,m_2,m_3=-\cn/2}^{\cn/2}|\vec{m}|=\int_{-\cn/2}^{\cn/2}\int_{-\cn/2}^{\cn/2}\int_{-\cn/2}^{\cn/2} \rmd m_1\rmd m_2\rmd m_3|\vec{m}|+O\left(\cn^{3}\right)\nonumber\\
&=&c \cn^4+O\left(\cn^{3}\right).
\ee
where $c$ is a numerical constant $c\simeq 0.480296$. Some details about the computation of $c$ can be found in Appendix \ref{Scaling of f(N)}. Finally, the one-loop effective action is given by
\be 
S_{1L}\left(\xi\right)&\simeq &4if(\cn)\xi+4\cn^3\log (\xi)\simeq c\cn^3\frac{4\pi i T}{\mu\sqrt{P_0}}-2\cn^3\log (P_0).\label{S1Laction}
\ee
If we use the relation $p=\frac{2\mu^{2}}{\beta a^{2}}P_{0}= j_0\frac{\ell_P^2}{a^2}$ with the area $j_0$ in Planck unit, we obtain
\be \label{eq:S1La}
S_{1L}(j_0)=i\cn^3\frac{T}{\sqrt{j_0}\ell_P}\chi_1-2\cn^3\log(j_0),\qquad \chi_1=\frac{4\sqrt{2}\pi c}{\sqrt{\b}}.
\ee
Notice that we have ignored a constant term in $S_{1L}$.

\subsection{Quantum equation of motion}\label{Quantum equations of motion}
In this subsection, we aim to compute the quantum equations of motion, namely the stationarity condition for the one-loop effective action determined by the critical point. This would shed light on the boundary state  ``preferred" by the one-loop effective action. In particular, the inclusion of the one loop functional determinant will furnish the information about the magnitude of the quantum correction coming from having integrated out the degrees of freedom.

At this stage, a remark is in order. Recall \eqref{LQGPI} - \eqref{hamiltonExp}. The coherent-state expectation value of the Hamiltonian in the action $S$ gives a semiclassical expansion in $t$ \cite{Giesel:2006um}:
\be
{\langle\psi_{g}^{t}\mid\hat{\mathbf{H}}\mid \psi_{g}^{t}\rangle}=\mathbf{H}[g]+t\mathbf{H}_1[g]+ O(t^2)
\ee
The leading order $\mathbf{H}[g]$ vanishes at the flat background, and its perturbation in $V^\rho$ contributes to the Hessian $H$ analyzed above. The next-to-leading order $\mathbf{H}_1[g]$ evaluated at the background contributes to the total effective action $\G$ at the same order in the expansion in $t$:
\be 
e^{i\G}=\exp\lt[i\G_0+O(t)\rt],\qquad i\G_0=S_{1L}-\frac{i\kappa}{a^2}T\mathbf{H}_1.
\ee
$\mathbf{H}_1$ being $\mathbf{H}_1[g]$ evaluated at the flat background has been computed in \cite{Zhang:2021qul}:
\begin{equation}
\kappa\mathbf{H}_1=\cn^3{H}_1\qquad
{H}_1=-\frac{3 a(1+\b^2) \left(1280 p^2-3072 p \coth (p)-1792 p \text{csch}(p)-5568\right)}{262144 \sqrt{\beta } p^{3/2}}.
\end{equation}
The total effective action $\G$ as a function of $p$ has the following leading order in $t$:
\be
\G_0(p)&=&\frac{\cn^3T}{a}w_{\rm eff}(p),\qquad \\
w_{\rm eff}(p)&=&\frac{4  \pi  \sqrt{2} c }{\sqrt{\beta } \sqrt{p}}+\frac{3  \left(\beta ^2+1\right) \left(1280 p^2-3072 p \coth (p)-1792 p \text{csch}(p)-5568\right)}{262144 \beta ^{3/2} p^{3/2}}+\frac{2i a}{T} \log (p).\nonumber
\ee
The imaginary part $\im(\G_0)\propto \log(p)$ is negligible for finite $p$ when $T$ is large. The effective action relates to the effective potential $V_{\rm eff}$ by $\G=\int\rmd^4x\sqrt{|\det(g)|}V_{\rm eff}$. This relation motivates us to express $\G_0$ as
\be
\G_0(p)&=&\frac{\cv_4}{a^4}V_{\rm eff}(p),\qquad \cv_4=\frac{a^3 \cn^3 T (\beta  p)^{3/2}}{2 \sqrt{2}},\\
V_{\rm eff}(p)&=&\frac{16 \pi  c}{\beta ^2 p^2}+\frac{3 \sqrt{2} \left(\beta ^2+1\right) \left(1280 p^2-3072 p \coth (p)-1792 p \text{csch}(p)-5568\right)}{131072 \beta ^3 p^3}+\frac{\left(4 \sqrt{2} i a\right) \log (p)}{\beta ^{3/2} p^{3/2} T}.\nonumber
\ee
Here $\cv_4$ is the total 4-volume from the metric \eqref{flatmetric111}. $V_{\rm eff}(p)$ is the effective potential of LQG. In contrast to the divergent effective potential in \eqref{Veffphi111}, $V_{\rm eff}(p)$ is finite if $p\neq 0$.

Indeed, $p$ needs to be determined by the quantum equation of motion ${\partial \G}/{\partial p}=0$, which gives
\be
\frac{3 \mathrm{csch}(p) \lt[12 \cosh (p)+14 p \coth (p)+24 p\, \mathrm{csch}(p)+7\rt]}{2048 \beta  p^{3/2}}&&\nonumber\\
+\frac{2 i a \sqrt{\beta }}{\left(\beta ^2+1\right) p T}-\frac{2 \sqrt{2} \pi  c}{\left(\beta ^2+1\right) p^{3/2}}
+\frac{3 \left(20 p^2+261\right)}{8192 \beta  p^{5/2}}&=&0
\ee
at the leading order in $t$. First we note that the imaginary part is suppressed for large $T$, rendering the solution real at leading order. Then, we can solve the equation numerically and find 2 solutions for $\b=1$
\begin{equation}
    p_*=j_0\ell_P^2/a^2 \simeq 0.08393\quad\text{and} \quad 288.90315
\end{equation}
Importantly, $p=0$ is not a solution.

Recall the discussion at the end of Section \ref{Effective action and coherent states}. Here $p_*$ as the analog of $\langle\theta^-|\hat{\phi}(x)|\theta^+\rangle$ uniquely fixes the boundary states by $p=p_*$ among coherent states corresponding to Minkowski spacetime. These coherent states lead to no solution to the quantum equation of motion unless $p=p_*$.

The resulting $p_*>0$ implies $j_0>0$ and the UV finiteness of the effective potential $V_{\rm eff}$. The exact value of $j_0$ still depend on the choice of length unit $a$, which originally comes from the definition of the coherent state $\psi^t_g$. If we choose $a=1\mathrm{mm}$, then $t=\ell_P^2/a^2\simeq 2.61\times 10^{-64}$, and we have $j_0\simeq 3.21394\times 10^{62} $ and $1.10629\times 10^{66}$ for these 2 solutions.

\section{Beyond the long wavelength approximation}\label{Beyond the long wavelength approximation}

In this section, we remove the long wavelength approximation that is proposed at the beginning of the last section. As a price, it is difficult to compute the determinant of Hessian analytically. Therefore, we proceed with the numerical computation. As we see in a moment, the resulting one-loop effective action gives the same formal expression as $S_{1L}$ in \eqref{S1Laction}, while the only difference is the numerical value of the constant $c$ and a numerical prefactor $y$ which takes into account the dependence on the momentum configuration.

The determinant of the Hessian matrix is given by 
\begin{align}
    \det(H) =\prod_{\vec{m} \in N^3} \prod_{k_0 \neq 0} DM(k_0,\vec{m})  =\prod_{\vec{m} \in L^3} \prod_{k_0=1}^{\infty} \left( DM(k_0,\vec{m}) \times DM(-k_0,\vec{m}) \right) 
\end{align}
Using the symmetry of $DM$ given in \eqref{eq:symmetry_DM}, we can express $\det (H)$ as 
\begin{align}
    \det (H) =\prod_{\vec{m} \in N^3} \prod_{k_0 \neq 0} DM(k_0,\vec{m})  =  \prod_{\vec{m} \in L^3, 0 \leq m_1 \leq m_2 \leq m_3} \left[\prod_{k_0=1}^{\infty} \left( DM(k_0,\vec{m}) \times DM(-k_0,\vec{m}) \right) \right]^{n_{\vec{m}}}, 
\end{align}
where $n_{\vec{m}}$ corresponds to the total number of permutations and parity transformations of $\vec{m}$. Now we start with the most general form of $DM$ \eqref{eq:general_DM}, we have 
\begin{align}
    DM(k_0, \vec{m}) DM(- k_0, \vec{m})=& \frac{\left(\frac{2 \pi}{T}k_0 \right)^{36}\mu^{72}} {a^{72} \beta^{36}} \prod_{i=1}^{6} \left(1 - \frac{\alpha_{i,\vec{m}}^2}{\left(\frac{2 \pi}{T}k_0 \right)^2 {P_0}} \right)\left(1 - \frac{\overline{\alpha_{i,\vec{m}}}^2 }{\left(\frac{2 \pi}{T}k_0 \right)^2 {P_0}} \right) . 
\end{align}
By fixing a lattice size $\cn$, $\alpha_{i,\vec{m}}$ are computed numerically for each $i$ and $\vec{m}$. {In practice, we can increase the lattice size up to $\cn=100$.} 

We use the product formula 
\begin{align}
    \prod_{k_0= 1}^{\infty} \left(1 - \frac{ \alpha^2 }{\left(\frac{2 \pi}{T}k_0 \right)^2 {P_0}} \right) = \frac{2 \sqrt{P_0} \sin\left( \frac{\alpha T}{2 \sqrt{P_0}} \right)}{\alpha T},
\end{align}
and a short-hand notation $z=T/(2\sqrt{P_0})$ to express $\det(H)$ as 
\begin{align}
    \det(H) = \mathcal{C} \prod_{\vec{m} \in L^3, m_1 \leq m_2 \leq m_3}  \left[ \prod_i^{6} \frac{ \sin\left( {\alpha_{i,\vec{m}} z } \right)}{\alpha_{i,\vec{m}} z } \frac{ \sin\left( {\overline{\alpha_{i,\vec{m}}} z } \right)}{\overline{\alpha_{i,\vec{m}}} z }  \right]^{n_{\vec{m}}}, 
\end{align}
where $\mathcal{C}$ is a field independent overall constant
\begin{align}
    \mathcal{C} =  \left[\prod_{k_0=1}^{\infty} \frac{\left(\frac{2 \pi}{T}k_0 \right)^{36}\mu^{72}} {a^{72} \beta^{36}} \right]^{\cn^3} .
\end{align}
The one-loop effective action is given by
\begin{align}\label{eq:S1Ln}
    S_{1L}(z) = - \log \det(H) = - \sum_{\vec{m} \in L^3, m_1 \leq m_2 \leq m_3} n_{\vec{m}} \sum_i^{6}  \log\lt[ \frac{ \sin\left( {\alpha_{i,\vec{m}} z } \right)}{\alpha_{i,\vec{m}} z } \frac{ \sin\left( {\overline{\alpha_{i,\vec{m}}} z } \right)}{\overline{\alpha_{i,\vec{m}}} z }  \rt]
\end{align}
For real $\alpha_{i,\vec{m}}$, there may be divergences when $  \alpha_{i,\vec{m}}z  \in \pi\mathbb{Z}$. The divergence can be removed by introducing the $i\epsilon$-regularization, namely  adding a phase to the time $T \to T e^{i\epsilon}$ with $\epsilon>0$. Consider the derivative of $S_{1L}$:
\be 
\frac{\mathrm{d} S_{1L}}{\mathrm{d} z} &=& -  \sum_{\vec{m} \in L^3, m_1 \leq m_2 \leq m_3} n_{\vec{m}} \sum_i^{6}  \left[- \frac{2}{z} + \alpha_{i,\vec{m}} \cot \left(  \alpha_{i,\vec{m}} z \right)+ \overline{\alpha_{i,\vec{m}}} \cot \left(  \overline{\alpha_{i,\vec{m}}} z \right) \right] \nonumber\\
&=&\frac{12 \cn^3}{z}  -  \sum_{\vec{m} \in L^3, m_1 \leq m_2 \leq m_3}  n_{\vec{m}} \sum_i^{6} \left[ \alpha_{i,\vec{m}} \cot \left(  \alpha_{i,\vec{m}} z \right)+ \overline{\alpha_{i,\vec{m}}} \cot \left(  \overline{\alpha_{i,\vec{m}}} z \right) \right].
\ee
As a difference from the long wavelength approximation, the entries of $\cot(\cdot) $ in $\rmd S_{1L}/\rmd P_0$ are complex since $\a_{i,\vec{m}}$ are generally complex. We are going to use the following two different expansions of $\cot(x)$, distinguishing whether $x\in\C$ belongs to the upper or lower half plane, in order to study the behavior as $T\to\infty$:
\begin{align}
    \cot(x) = \begin{cases}
    -i -2 i \sum_{k=1}^{\infty} \exp(2 i k x), & \text{when}\  \Im(x)>0,\\
    i + 2 i \sum_{k=1}^{\infty} \exp(- 2 i k x)& \text{when}\  \Im(x)<0.
    \end{cases}
\end{align}
The sum in $\rmd S_{1L}/\rmd z$ contains the terms of four different types:

\begin{itemize}

\item When $\Im (\alpha_{i,\vec{m}}) > 0$, we use the following expansion
\be 
&&{\alpha_{i,\vec{m}} \cot \left(  {\alpha_{i,\vec{m}} z } \right)}+  \overline{\alpha_{i,\vec{m}}} \cot \left( {\overline{\alpha_{i,\vec{m}}} z } \right)\nonumber\\
    &=& -{\alpha_{i,\vec{m}} } \left[ i + 2 i \sum_{k=1}^{\infty} \exp \left( {2 i k \alpha_{i,\vec{m}} z  }  \right) \right] + {\overline{\alpha_{i,\vec{m}}}  }\left[i + 2 i \sum_{k=1}^{\infty} \exp \left( - {2 i k\overline{\alpha_{i,\vec{m}}} z} \right) \right] .
\ee
Every exponential is suppressed as $z=T/(2\sqrt{P_0})\to\infty$. After neglecting the exponentially suppressed terms, we obtain
\be
{\alpha_{i,\vec{m}} \cot \left(  {\alpha_{i,\vec{m}} z } \right)}+  \overline{\alpha_{i,\vec{m}}} \cot \left( {\overline{\alpha_{i,\vec{m}}} z } \right)\simeq  2 \Im(\alpha_{i,\vec{m}})  .
\ee

\item When $\Im (\alpha_{i,\vec{m}}) < 0$,
\be
&&{\alpha_{i,\vec{m}} \cot \left(  {\alpha_{i,\vec{m}} z } \right)}+  \overline{\alpha_{i,\vec{m}}} \cot \left( {\overline{\alpha_{i,\vec{m}}} z } \right)\nonumber\\
    &=& {\alpha_{i,\vec{m}} } \left[ i + 2 i \sum_{k=1}^{\infty} \exp \left( {2 i k \alpha_{i,\vec{m}} z  }  \right) \right] - {\overline{\alpha_{i,\vec{m}}}  }\left[i + 2 i \sum_{k=1}^{\infty} \exp \left( - {2 i k\overline{\alpha_{i,\vec{m}}} z} \right) \right] \nonumber\\
    &\simeq& -2\Im(\alpha_{i,\vec{m}}) .
\ee

\item When $\Im (\alpha_{i,\vec{m}}) =0$, due to the $i\epsilon$-regularization
\be 
2{\alpha_{i,\vec{m}} \cot \left(  {\alpha_{i,\vec{m}} z e^{i\epsilon}} \right)}
&=& -2 \alpha_{i,\vec{m}} \left[ i + 2 i \sum_{k=1}^{\infty} \exp \left( 2 i k \alpha_{i,\vec{m}} z  \right) \right]\nonumber\\
&\simeq &-2 i \alpha_{i,\vec{m}}.
\ee

\item When $\a_{i,\vec{m}}=0$,
\be
{\alpha_{i,\vec{m}} \cot \left(  {\alpha_{i,\vec{m}} z } \right)}+  \overline{\alpha_{i,\vec{m}}} \cot \left( {\overline{\alpha_{i,\vec{m}}} z } \right)=\frac{2}{z}.
\ee

\end{itemize}

\begin{figure}
    \centering
    \includegraphics[width=0.45\linewidth]{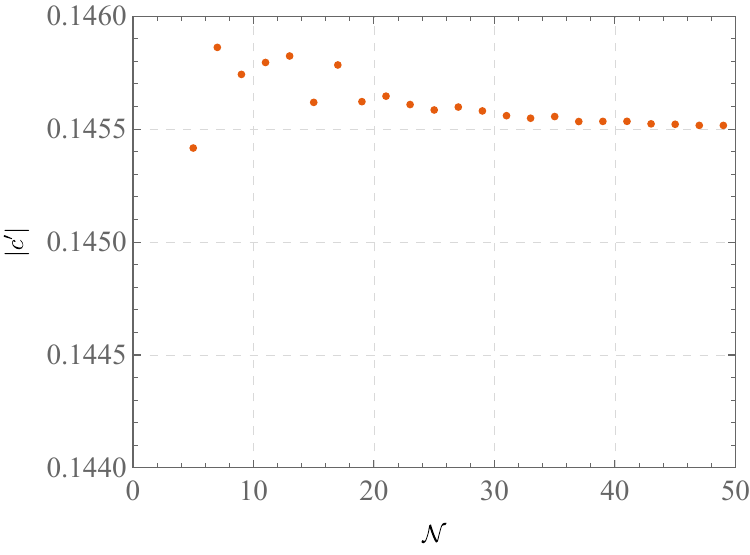}
    \includegraphics[width=0.45\linewidth]{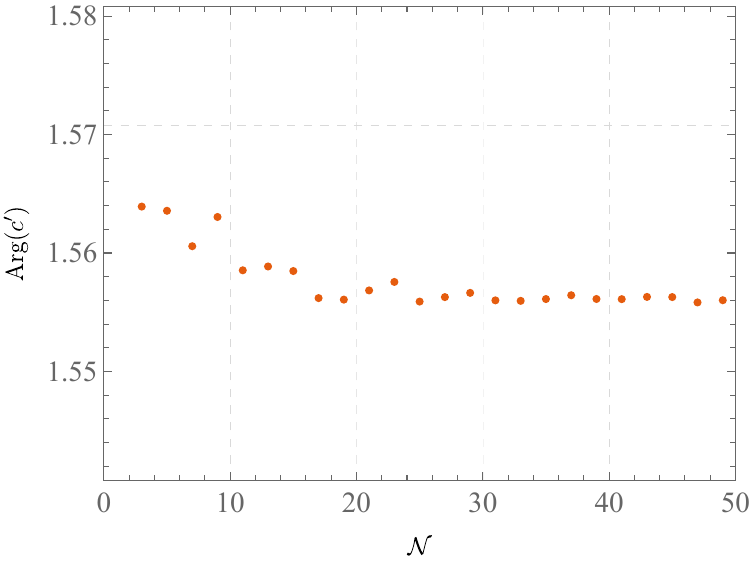}
    \includegraphics[width=0.45\linewidth]{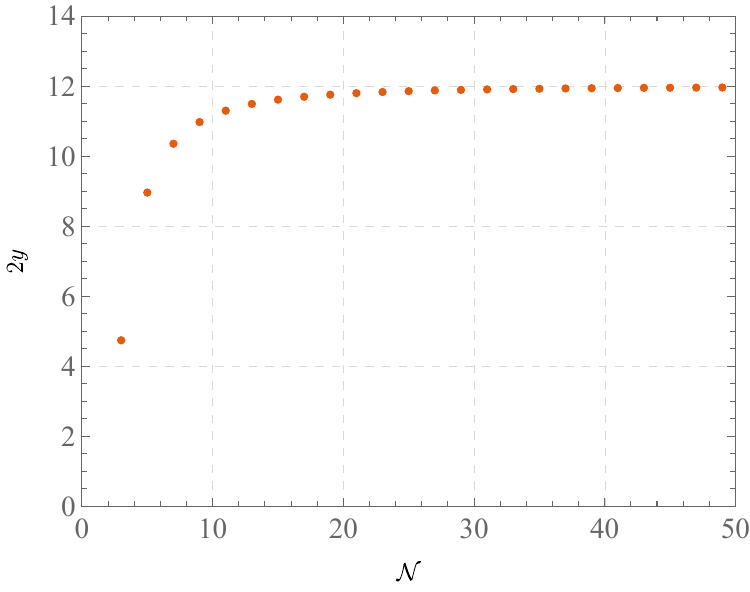}
    \includegraphics[width=0.45\linewidth]{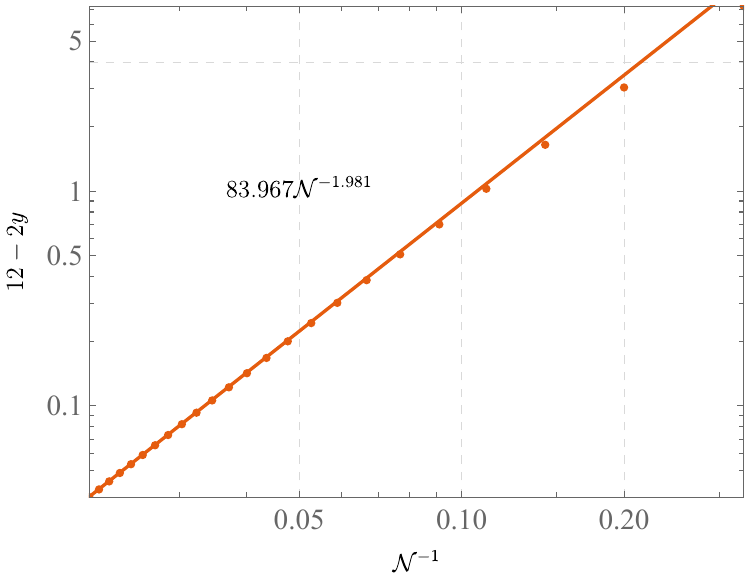}
    \caption{The plots 
 of $c'$ (top) and $2y$ (bottom) versus lattice size $\mathcal{N}$. The argument of $c'$  goes slight below $\frac{\pi}{2}$. $2y$ is increasing and approaches its maximum value $12$. }
    \label{fig:cp_2y}
\end{figure}

As a result, altogether we have
\be\label{eq:ds_regular}
    \frac{\mathrm{d} S_{1L}}{\mathrm{d} z} &=&\frac{{ 2y} \cn^3}{z}  +\frac{\cn^3}{\mu} 8\pi ic',\\
    8\pi ic'&=&-\frac{\mu}{\cn^3} \sum_{\vec{m} \in L^3, m_1 \leq m_2 \leq m_3} n_{\vec m}  \sum_i^{6} \left(- 2i \alpha_{i,\vec{m}} \delta_{\Im \alpha_{i,\vec{m}}, 0} +  {2|\Im{\alpha_{i,\vec{m}}} | } \right).
\ee
{Generally, $2y<12$ is due to the vanishing $\a_{i,m}$ for some $i$ and $\vec{m}$}. Integrating this result gives the expression of $S_{1L}$ up to a constant term
\be\label{S1LN}
S_{1L}=c'\cn^3 \frac{4 \pi i T}{\mu  \sqrt{P_0}}- y \cn^3 \log (P_0)\;.
\ee
This result is formally the same as the result \eqref{S1Laction} from the long wavelength approximation, up the difference of the coefficients $c'$ and $y$. The behavior of $c'$ and $y$ with lattice size $\mathcal{N}$ is given in Fig. \ref{fig:cp_2y}. 
{From Fig. \ref{fig:cp_2y}, both $c'$ and $y$ converge to a constant value for sufficiently large $\mathcal{N}$. The argument of $c'$ converges to a value slightly below $\frac{\pi}{2}$, namely its real part is small compared to the imaginary part. Since the long-wavelength approximation gives only a real $c$, this implies that the momentum cutoff will introduce a large contribution to the effective action and the quantum equation of motion, leading to a complex critical point. Meanwhile, $2y$ will converge to its maximum value $12$ with the difference scaled approximately as $\mathcal{N}^{-2}$ if $\mathcal{N}$ is sufficiently large. The reason is that the vanishing $\alpha_{i,m}$ appear only for special configurations, e.g. $\vec{m}=(m,0,0)$. So the total number of vanishing $\alpha_{i,m}$ will scale with $\mathcal{N}$ instead of $\cn^3$.  }

\section{Conclusion}\label{sec:8}
In this work we  presented a novel method that combines path integral effective techniques with the canonical formulation of LQG, by means of  the recently introduced coherent state path integral formulation of LQG \cite{Han:2019vpw}.  By utilizing coherent states, a novel discrete reduced phase path integral formulation for the transition amplitude governed by the physical Hamiltonian  in the context of gravity coupled to dust has been derived \cite{Han:2020chr}. Building on this, we calculate the corresponding one-loop effective action and the quantum equation of motion.

We review the standard QFT computation of the one-loop effective action for Einstein-Hilbert gravity and we emphasize which  UV divergences arise. In contrast to these procedures, LQG provides a  natural UV cutoff given by the minimal eigenvalue of $j$. By a power counting argument, we argue how no power-law divergences should arise in transition amplitudes. On top of that,  the cutoff $j_0$ plays another role, as part of the gravitational dynamics of the system, which is selected imposing the quantum equation of motion.

The computation of the one-loop effective action is a powerful tool to integrate out the degrees of freedom and to encode the quantum correction into a low-energy description. We compute it for the coherent state path integral, emphasizing the difference with the standard prescription based on asymptotic states.
In particular, we keep track of the choice of the boundary state in deriving the generating functional and the effective action, back to its role as an expectation value over quantum states. On the other hand, seeking for the on-shell solution, the  existence of it furnishes  the
constraint to the boundary states: since the effective action is a function of the boundary states, the requirement of the existence of an on-shell solution selects the boundary states as the critical point of the effective action.

Considering a finite cubic lattice discretizing the spatial slice, we choose a flat background and perturb the connection and the gauge-covariant flux variables around that background. 
This allows to compute the $\log \det (H)$ of the Hessian of the perturbations.
Furthermore, working on a lattice on a flat background, we can work in Fourier space.

As an intermediate step, we compute the propagator of the perturbations integrating out the holonomies perturbations. The eigenvalue problem can  be solved numerically, selecting the contributing modes and their scaling, depending on the different configurations.  In the short wavelength approximation,   a separation between gauge and physical degrees of freedom is not possible. We perform several numerical studies for the convergence of the eigenvalues and the poles to the continuum limit increasing the lattice points $\mathcal{N}$. On the other hand, we also investigate numerically the scaling of the poles for different configurations, when the momentum $\vec{k}$ scales with $\mathcal{N}$. We derive the propagator in the continuum limit, corresponding to the linearized equation of motion, including the linearized Gauss constraint. We recognize the vector, tensor and scalar modes and their eigenvectors and eigenvalues. We emphasize that our path integral is describing only the dynamics of the graviton, two (tensor) physical degrees of freedom, being defined on the reduced phase space. 

In QFT, evaluating the functional determinant using the path integral corresponds to performing a Gaussian integral over the fluctuations. In contrast, within a discretized setting, this process requires diagonalizing the Hessian and calculating the product of its eigenvalues. We compute analytically the one-loop effective action in the long-wavelength approximation and study its behavior numerically beyond such approximation. We compare with the expression of the effective action from the dimensional argument and compute the quantum equation of motion.
Within the long-wavelength approximation, we determine the one-loop contribution to the effective action, which exhibits at most a logarithmic dependence  on $j_0$. This contrasts with standard QFT computations, where fine-tuning procedures are required to eliminate power-law divergences in the cutoff. 
Moreover, our procedure did not require introducing any unphysical cutoff scale. This is a significant advantage when working with a lattice-regularized theory like LQG, where a natural cutoff is provided by the minimal eigenvalue of $j$. Furthermore, we use  the one-loop effective action to derive the quantum equation of motion. This approach helps identifying the boundary state that aligns with quantum corrections and ensures the validity of the long-wavelength approximation.
Beyond this regime, we can numerically evaluate in full generality  the expression of the one-loop determinant. Having the same structural form as in the continuum limit, we compute numerically the coefficients in this regime for different momentum configurations. The one-loop effective action is given by \eqref{eq:S1La} and \eqref{S1LN} in the long and short wavelength approximation, respectively.

\bigskip

This first calculation of the one-loop effective action within the LQG coherent state path integral lays the groundwork for future research and validates the endeavor for applying QFT techniques within the canonical formulation of quantum gravity. Ideally, one would try to make the formal connection with the effective description \cite{Bojowald:2011aa,Zhang:2021xoa,Zhang:2024khj}.

From a QFT point of view, at this stage, it would be interesting to make contact with the renormalization program, constructing a beta function to describe the running of the coupling constants. Ideally, one could then undo the choice of a specific background, seeking for the background independence realized in Asymptotic Safety.
As a first step, this implies that the theory should be covariantized to enable a comparison of operators for renormalization \cite{Ferrero:2024rvi}. A natural example would be the mimetic gravity framework, which has been introduced within the context of canonical quantum gravity using dust as a clock \cite{Han:2022rsx}.

An important feature of our computation is that the analysis is performed entirely in Lorentzian signature. Throughout, we highlight the significance of the boundary state dependence. 
It would be intriguing to apply our newly developed methods to gain insights into the nature of the preferred gravitational vacuum state \cite{Ashtekar:1993wf,Ashtekar:1994mh} and the influence on it due to the discretization \cite{Dittrich:2012jq, Dittrich:2013xwa, Dittrich:2014wpa}. Additionally, it would be an interesting benchmark to compare our effective action  with the one derived from  effective spinfoams for 4D Lorentzian quantum gravity \cite{Asante:2021zzh,Borissova:2022clg}.

Finally, as a natural extension, our analysis could be applied to a cosmological background, potentially with flat spatial slices, in order to  access to the spatial Fourier space. This approach would enable the derivation of  one-loop quantum corrections to the cosmological effective action \cite{Cheung:2007st,Weinberg:2008hq,Baumann:2010tm,Battye:2012eu} and the consequences on the power spectrum \cite{Giesel:2007wi, Agullo:2015tca,Gomar:2015oea}.

\section*{Acknowledgments}

M.H. receives supports from the National Science Foundation through grant PHY-2207763, the College of Science Research Fellowship at Florida Atlantic University, and the Blaumann Foundation. R.F. and H.L. are grateful for the hospitality of Perimeter Institute where part of this work was carried out. Research
at Perimeter Institute is supported in part by the Government of Canada through the Department
of Innovation, Science and Economic Development and by the Province of Ontario through the
Ministry of Colleges and Universities. This work was supported by a grant from the Simons
Foundation (1034867, Dittrich).

\appendix

\section{Scaling of $f(\cn)$}\label{Scaling of f(N)}

We apply the following Euler–Maclaurin formula to compute $f(\cn)=\sum_{{m}_1,m_2,m_3=-\cn}^\cn|\vec{m}|$
\be 
\sum_{i=m}^n f(i)=\int_m^n f(x) d x+\frac{f(n)+f(m)}{2}+\sum_{k=1}^{\left\lfloor\frac{p}{2}\right\rfloor} \frac{B_{2 k}}{(2 k)!}\left(f^{(2 k-1)}(n)-f^{(2 k-1)}(m)\right)+R_p,
\ee
where $B_{2k}$ is the $k$th Bernoulli number. 

Since we have 
\be 
\left[\partial_{m_{i}}^{n}\sqrt{m_{1}^{2}+m_{2}^{2}+m_{3}^{2}}\right]_{m_{i}\to\pm \cn}\leq  O(\cn),\qquad n=0,1,2,\cdots ,
\ee
we obtain that for the subleading terms in the Euler–Maclaurin formula
\be 
\sum_{m_{j}\neq m_{i}}\left[\partial_{m_{i}}^{2k-1}\sqrt{m_{1}^{2}+m_{2}^{2}+m_{3}^{2}}\right]_{m_{i}\to\pm \cn}\leq O\left(\cn^{3}\right),\\
\sum_{m_{l}}\int_{-\cn}^\cn dm_{j}\left[\partial_{m_{i}}^{2k-1}\sqrt{m_{1}^{2}+m_{2}^{2}+m_{3}^{2}}\right]_{m_{i}\to\pm \cn}\leq O\left(\cn^{3}\right),\\
\int_{-\cn}^\cn dm_{j}\int_{-\cn}^\cn dm_{l}\left[\partial_{m_{i}}^{2k-1}\sqrt{m_{1}^{2}+m_{2}^{2}+m_{3}^{2}}\right]_{m_{i}\to\pm \cn}\leq O\left(\cn^{3}\right).
\ee
which implies that 
\be 
\sum_{{m}_1,m_2,m_3=-\cn}^\cn|\vec{m}|=\int_{-\cn}^\cn\int_{-\cn}^\cn\int_{-\cn}^\cn \rmd m_1\rmd m_2\rmd m_3|\vec{m}|+O\left(\cn^{3}\right).
\ee
Note that this will result in a scaling as $\mathcal{N}^4$, which is consistent with the results found in \cite{Becker:2021pwo,Ferrero:2024yvw}.

\bibliographystyle{jhep}
\bibliography{ref}
	
\end{document}